\documentclass[twocolumn]{aastex631}

\usepackage{amsmath}	
\usepackage[percent]{overpic}

\shorttitle{Trends in chemistry and dynamics in ultra-hot Jupiters}
\shortauthors{Gandhi et al.}

\begin{document}

\title{Retrieval survey of metals in six ultra-hot Jupiters: Trends in chemistry, rain-out, ionisation and atmospheric dynamics}

\correspondingauthor{Siddharth Gandhi}
\email{gandhi@strw.leidenuniv.nl}

\author[0000-0001-9552-3709]{Siddharth Gandhi}
\affiliation{Leiden Observatory, Leiden University, Postbus 9513, 2300 RA Leiden, The Netherlands}
\affiliation{Department of Physics, University of Warwick, Coventry CV4 7AL, UK}
\affiliation{Centre for Exoplanets and Habitability, University of Warwick, Gibbet Hill Road, Coventry CV4 7AL, UK}

\author[0000-0002-3239-5989]{Aurora Kesseli}
\affiliation{IPAC, Mail Code 100-22, Caltech, 1200 E. California Blvd., Pasadena, CA 91125, USA}

\author{Yapeng Zhang}
\affiliation{Leiden Observatory, Leiden University, Postbus 9513, 2300 RA Leiden, The Netherlands}

\author{Amy Louca}
\affiliation{Leiden Observatory, Leiden University, Postbus 9513, 2300 RA Leiden, The Netherlands}

\author{Ignas Snellen}
\affiliation{Leiden Observatory, Leiden University, Postbus 9513, 2300 RA Leiden, The Netherlands}

\author{Matteo Brogi}
\affiliation{Department of Physics, University of Warwick, Coventry CV4 7AL, UK}
\affiliation{INAF-Osservatorio Astrofisico di Torino, Via Osservatorio 20, I-10025 Pino Torinese, Italy}
\affiliation{Dipartimento di Fisica, Universit\`a degli Studi di Torino, via Pietro Giuria 1, I-10125, Torino, Italy}

\author[0000-0002-0747-8862]{Yamila Miguel}
\affiliation{Leiden Observatory, Leiden University, Postbus 9513, 2300 RA Leiden, The Netherlands}
\affiliation{SRON Netherlands Institute for Space Research, Niels Bohrweg 4, 2333 CA, Leiden, the Netherlands}

\author[0000-0002-2891-8222]{Núria Casasayas-Barris} 
\affiliation{Instituto de Astrofísica de Canarias, Vía Láctea s/n, E-38205 La Laguna, Tenerife, Spain}
\affiliation{Departamento de Astrofísica, Universidad de La Laguna, Spain}

\author[0000-0002-8573-805X]{Stefan Pelletier}
\affiliation{Department of Physics and Trottier Institute for Research on Exoplanets, Universit\'{e} de Montr\'{e}al,
1375 Avenue Th\'{e}r\`{e}se-Lavoie-Roux, Montreal, H2V 0B3, Canada}

\author{Rico Landman}
\affiliation{Leiden Observatory, Leiden University, Postbus 9513, 2300 RA Leiden, The Netherlands}

\author{Cathal Maguire}
\affiliation{School of Physics, Trinity College Dublin, University of Dublin, Dublin 2, Ireland}

\author
{Neale P. Gibson}
\affiliation{School of Physics, Trinity College Dublin, University of Dublin, Dublin 2, Ireland}



\begin{abstract}
Ground-based high-resolution spectroscopy (HRS) has detected numerous chemical species and atmospheric dynamics in exoplanets, most notably ultra-hot Jupiters (UHJs). However, quantitative estimates on abundances have been challenging but are essential for accurate comparative characterisation and to determine formation scenarios. In this work we retrieve the atmospheres of six UHJs (WASP-76~b, MASCARA-4~b, MASCARA-2~b, WASP-121~b, HAT-P-70~b and WASP-189~b) with ESPRESSO and HARPS-N/HARPS observations, exploring trends in eleven neutral species and dynamics. While Fe abundances agree well with stellar values, Mg, Ni, Cr, Mn and V show more variation, highlighting the difficulty in using a single species as a proxy for metallicity. We find that Ca, Na, Ti and TiO are under-abundant, potentially due to ionisation and/or night-side rain-out. Our retrievals also show that relative abundances between species are more robust, consistent with previous works. We perform spatially- and phase-resolved retrievals for WASP-76~b and WASP-121~b given their high signal-to-noise observations, and find the chemical abundances in each of the terminator regions are broadly consistent. We additionally constrain dynamics for our sample through Doppler shifts and broadening of the planetary signals during the primary eclipse, with median blue shifts between $\sim$0.9-9.0~km/s due to day-night winds. Furthermore, we constrain spectroscopic masses for MASCARA-2~b and HAT-P-70~b consistent with their known upper limits, but we note that these may be biased due to degeneracies. This work highlights the importance of future HRS studies to further probe differences and trends between exoplanets.
\end{abstract}


\section{Introduction} \label{sec:intro}

We are entering the era of comparative exoplanet atmosphere science, with a new generation of observatories probing the atmospheres of planets in ever increasing detail to determine the similarities of these worlds to each other and our own Solar System. Recently, ground-based facilities capable of observing exoplanets at very high spectral resolution (R$\gtrsim$25,000) \citep[e.g.,][]{mayor2003, jurgenson2016, pepe2021} have allowed for unprecedented detail into the chemical inventory of exoplanets, most notably for ultra-hot Jupiters with temperatures in excess of 2000~K. These strongly irradiated exoplanets are unique laboratories which exhibit a wide range of chemistry and physics \citep[e.g.,][]{fortney2008, parmentier2018}, and are excellent targets to study planetary chemical composition, particularly of refractory species. The high resolution observations allow us to to resolve many thousands of individual absorption lines in the planetary spectrum \citep[see e.g.,][]{birkby2018}. The difference in the position and relative strength of the spectral lines between species allows us to unambiguously detect a range of atoms, molecules and ions in the atmosphere. A wide range of chemical species have been observed in UHJs \citep[e.g.,][]{casasayas-barris2018, seidel2019, hoeijmakers2019, ben-yami2020, cont2021, merritt2021, kesseli2022, vansluijs2023}. In particular, gaseous Fe has been shown to be prevalent in a number of UHJs, under both primary and secondary eclipse observations \citep[e.g.,][]{hoeijmakers2018, nugroho2020, hoeijmakers2020, pino2020, yan2020, casasayas-barris2022, kasper2022}.

High resolution observations of UHJs have also allowed us to explore winds present in their atmospheres. Global winds result in a Doppler shift and/or broadening of the overall spectrum due to the motion of the constituents of the planetary atmosphere towards or away from the observer, and can most clearly be constrained during primary eclipse \citep[e.g.,][]{snellen2010, louden2015, brogi2016}. These transit observations have shown that wind speeds of $\sim$km/s are present in the upper atmospheres of these highly irradiated planets \citep[e.g.,][]{seidel2021, langeveld2022}, and allowed for theoretical studies of the circulation patterns \citep{tan2019}. In addition, we have been able to explore variations of the spectra during the transit \citep{ehrenreich2020, kesseli2021}, and have shown that condensation, cloud formation on one side of the planet and thermal asymmetries can result in shifts in the strength and position of spectral features as the transit progresses \citep[e.g.,][]{savel2022, gandhi2022}. The mechanisms that drive the speed and spatial distribution of the winds are influenced by atmospheric drag \citep[e.g.,][]{wardenier2021} and by magnetic fields \citep[e.g.,][]{beltz2022}. Atmospheric winds are also coupled to the formation of clouds in the atmosphere \citep{komacek2022}. Hence the dynamics are important quantities for understanding the various processes within these planets as well as obtaining reliable chemical compositions.

Recently, retrievals of high resolution spectra have given us the ability to quantitatively constrain on the chemical abundances of gaseous species in the atmospheres of hot and ultra-hot Jupiters \citep{brogi2017, gandhi2019_hydrah, pelletier2021,line2021, gibson2022}. Such retrievals have become possible thanks to developments in likelihood based approaches for high resolution spectra \citep{brogi2017, brogi2019, gibson2020}, and often explore many millions of models over a wide range in parameter space to fit the observations. The high resolution modelling required for the retrievals requires significant computational power, but allows for quantitative estimates to be placed on the atmospheric constituents. The abundance constraints are key to determine the refractory and volatile content of the atmosphere and thus understand how these planets formed and migrated to their current location close to their host star \citep[e.g.,][]{madhu2014, lothringer2021, knierim2022}.

Given the numerous recent high precision observations of UHJ atmospheres, a key step is to be able to compare and contrast the chemical composition of these exoplanets in order to observe trends as well as to explore potential formation pathways. In this work we perform high resolution retrievals of terminator observations of six UHJs, and constrain eleven neutral atomic and molecular species expected to be prevalent in their atmospheres. Furthermore, we constrain the winds on the terminator of these UHJs through velocity shifts of the planetary signal and compare the trends across our sample. We use optical observations obtained with the ESPRESSO and HARPS-N/HARPS spectrographs as shown in Table~\ref{tab:planet_params}. Our sampl of UHJs were chosen given their similar $\sim$2,200-2,600~K equilibrium temperatures and the multiple species which are prominent in their atmosphere \citep[e.g.,][]{cabot2020, merritt2021, tabernero2021, kesseli2022}. 

\begin{table*}[]
    \centering
\begin{tabular}{c|cccccc}
Parameter &\bf{WASP-76~b} & \bf{MASCARA-4~b} & \bf{MASCARA-2~b}$^{\dagger}$ & \bf{WASP-121~b} & \bf{HAT-P-70~b} & \bf{WASP-189~b} \\
\hline
Reference              & (1) & (2) & (3) & (4) & (5) & (6) \\
Instrument & ESPRESSO & ESPRESSO & HARPS-N & ESPRESSO & HARPS-N & HARPS-N/S \\
\hline
Obs. Dates & 2018-09-02 & 2020-02-12 & 2017-08-16 & 2019-01-06 & 2020-12-18 & 2019-04-14 \\
 & 2018-10-30 & 2020-02-29 & 2018-07-12 & & & 2019-04-25 \\
 & & & 2018-07-19 & & & 2019-05-06 \\
 & & & & & & 2019-05-14\\
\hline
T$_\mathrm{eq}$/K & 2210 & 2250 & 2260$^{10}$ & 2358$^{13}$ & 2562$^{15}$ & 2641$^{16}$\\
R$_*$/R$_\odot$ & 1.756 & 1.79$^8$ & 1.60$^{10}$ & 1.44 & 1.858$^{15}$ & 2.36$^{17}$ \\
$[$Fe/H$]$           & 0.366 & $\sim$0.0$^9$ & -0.02$^{10}$ & 0.13 & -0.059$^{15}$ & 0.29$^{17}$  \\
R$_\mathrm{p}$/R$_\mathrm{J}$   & 1.854 & 1.515 & 1.83$^{10}$ & 1.753$^{14}$ & 1.87$^{15}$ & 1.619$^{17}$\\
M$_\mathrm{p}$/M$_\mathrm{J}$     & 0.894 & 1.675 & $<$3.51$^{11}$ & 1.157$^{14}$ & $<$6.78$^{15}$ & 1.99$^{17}$\\
T$_c$/BJD$_\mathrm{TDB}$               & 2458080.6257$^7$ & 2458909.66419 & 2458312.58564$^7$ & 2457705.37012$^7$ & 2459175.05277$^7$ & 2458926.5416960$^{17}$ \\
T$_c$ 1$\sigma$ error  & 0.00013$^7$ & 0.00046 & 0.00016$^7$ & 0.00012$^7$ & 0.00017$^7$ & 0.000065$^{17}$ \\
P/days                  & 1.80988058$^7$ & 2.8240932 & 3.47410024$^7$ & 1.27492483$^7$ & 2.74432$^7$ & 2.7240308$^7$ \\
P 1$\sigma$ error         & 1.8E-07$^7$ & 4.6E-06 &  6.1E-07$^7$ & 1.1E-07$^7$ & 1E-06$^7$ & 4.2E-06$^{17}$ \\
K$_*$/ km\,s$^{-1}$       & 0.1156 & 0.1659 & 0.0 & 0.177$^{14}$ & 0.0 & 0.182$^{17}$ \\
K$_\mathrm{p}$/ km\,s$^{-1}$  & 196.5$\pm1$ & 182$\pm5$ & 169$\pm6^{12}$ & 221$\pm4.5$ & 187$\pm$4 & 200.7$\pm$4.9 \\
V$_\mathrm{sys}$/km\,s$^{-1}$             & $-1.16\pm0.0025$ & $–5.68\pm$0.09 & $-24.48\pm0.04^{12}$ & 38.198$\pm0.002$ & 25.26$\pm0.11^{15}$ & $-24.452\pm0.12^{17}$\\
\end{tabular}
    \caption{Observations and planetary parameters used for our retrievals. The references for the sources of the data are as follows (1) \citet{ehrenreich2020} (Programme: ESO 1102.C-0744), (2) \citet{zhang2022} (ESO 0104.C-0605), (3) \citet{casasayas-barris2019} (PID CAT17A-38), (4) \citet{borsa2021} (ESO 1102.C-0744), (5) \citet{bello-arufe2022} (A42TAC27), (6) \citealt{prinoth2022} (ESO 0103.C-0472, CAT19A-97). Unless otherwise specified, the references for the parameters are taken from the data references. For T$_c$ and P, we used the most precise transit timing measurements, many of which are from TESS, (7) \citet{ivshina2022}.  Other references are (8) \citet{ahlers2020}, (9) \citet{dorval2020}, (10) \citet{talens2018}, (11) \citet{lund2017}, (12) \citet{rainer2021}, (13) \citet{delrez2016}, (14) \citet{bourrier2020}, (15) \citet{zhou2019}, (16) \citet{anderson2018}, and (17) \citet{lendl2020}. For planets where only an upper mass limit was detected, we used a K$_{*}$ value of 0.0. $^{\dagger}$ MASCARA-2~b is also known as KELT-20~b.}
    \label{tab:planet_params}
\end{table*}

We retrieve the chemical species in the atmosphere using the HyDRA-H high resolution retrieval models \citep{gandhi2019_hydrah}. We perform retrievals with chemical abundances of eleven species as free parameters: Fe, Mg, Ni, Cr, Mn, V, Ca, Ti, TiO, TiH and Na. For this work we do not retrieve ionic species in order to keep the retrievals computationally tractable and given that ionic species generally have more complex vertical chemical abundance profiles. For two of the planets in our survey, WASP-76~b and WASP-121~b, we perform additional phase-separated and spatially-resolved retrievals with HyDRA-2D given the high signal-to-noise and high resolution R=140,000 ESPRESSO observations available \citep{gandhi2022}. This allows us to further explore the variation of the chemical species and dynamics in the different regions of the terminator for these two planets. 

The next section discusses the retrieval setup and data analysis, followed by the results and discussion where we explore the constraints on the atmospheric chemistry and dynamics. Finally, we present the concluding remarks in section~\ref{sec:conc}.

\section{Methods}\label{sec:methods}

In this section we discuss the modelling setup and data analysis. We use HyDRA-H and HyDRA-2D to perform the retrievals of the exoplanets in our survey \citep{gandhi2019_hydrah, gandhi2022}. These models explore a wide range of compositions, temperature profiles, cloud/opacity decks and wind speed distributions to characterise the atmosphere from ground-based high-resolution observations. We use primary eclipse observations with the ESPRESSO spectrograph on the VLT \citep{pepe2021}, HARPS-N spectrograph on the TNG and HARPS on La Silla \citep{mayor2003}, which cover the optical wavelength range ($\sim$0.4-0.8~$\mu$m). Table~\ref{tab:planet_params} shows the system parameters for each of the planets in our survey and the observations we used. We discuss the retrieval setup, our target sample and data analysis in further detail below.

\subsection{Atmospheric modelling and retrieval setup}

We perform our retrievals for the six planets using HyDRA-H \citep{gandhi2019_hydrah}. Our setup assumes free parameters for the atmospheric chemistry (volume mixing ratio for each chemical species) and a vertically varying temperature profile. Our atmosphere is assumed to be H/He-rich, which is a valid assumption for these UHJs given the high temperatures will lead to substantial dissociation of H$_2$ in the photosphere \citep[see e.g.,][]{parmentier2018, gandhi2022}.

To generate our model spectra for each of the exoplanets we use the latest and most precise line lists to determine the line-by-line opacity from each of the spectrally relevant species in the atmosphere. The opacity for the atomic species is determined using the Kurucz line list \citep{kurucz1995}. The cross sections for the molecular species TiO and TiH are calculated using the TOTO \citep{tennyson2016, mckemmish2019} and MoLLIST \citep{burrows2005, bernath2020} line lists respectively. These high-temperature line lists are the most suitable to the high resolution observations in our study given their accurately determined line positions. For each spectral line we determine the broadening due to the temperature as well as pressure, which results in a Voigt line profile. The pressure broadening coefficients are calculated using the method of \citet{sharp2007} and described in further detail in \citet{gandhi2022} for atomic species, and \citet{gandhi2020_cs} for the molecules. For each species we determine the total cross section by summing the contributions from each of the lines, calculated between 0.33-50~$\mu$m with a wavenumber spacing of 0.01~cm$^{-1}$. This corresponds to a spectral resolution of R=$2.5\times10^6$ at 0.5~$\mu$m.

We additionally include the opacity/haze deck of the atmosphere into our retrieval as an additional source of absorption, which originates from any potential cloud condensation or haze \citep[e.g.,][]{gao2021, komacek2022} as well as other species with strong continuum-like opacity such as TiO or H- \citep[e.g,][]{hubeny2003, arcangeli2018, parmentier2018, gandhi2020_h-}. However, our high resolution observations are generally more sensitive to higher altitudes above the regions with significant opacity \citep[see e.g.,][]{gandhi2020_clouds, hood2020}, but clouds and hazes can impact the spectra and hence we include them into our retrieval to avoid any biases in the chemical abundances.

For each of the spectral models in our retrieval we generate transmission spectra using numerical radiative transfer \citep[e.g.,][]{pinhas2018}. We determine the temperature profile using the parametrisation of \citet{madhu2009}, which uses 6 free parameters, and allows for non-inverted, isothermal and inverted profiles. We also include free parameters $\Delta K_\mathrm{p}$, $\Delta V_\mathrm{sys}$ and $\delta V_\mathrm{wind}$ into our retrieval, which are the deviation from the planet's known orbital velocity, deviation from the known systemic velocity and the full-width half-maximum (FWHM) of the wind profile of the atmosphere \citep[see][for further details]{gandhi2022}. For each model we generate spectra at a resolution of R=500,000 in the wavelength range of the observations, and convolve to the instrumental resolution of R=140,000 for ESPRESSO and R=115,000 for HARPS-N/HARPS.

We perform additional retrievals for WASP-76~b and WASP-121~b with HyDRA-2D \citep{gandhi2022}, separating the regions of the terminator into a morning (leading) limb and an evening (trailing) limb, with separate chemical abundances, temperature profile and opacity deck for each half of the terminator. This model has been demonstrated on the WASP-76 ESPRESSO observations, showing differing chemical, thermal and opacity deck constraints between the two halves of the terminator. Previous work has shown that multi-dimensional retrievals are important for such hot exoplanets \citep[e.g.,][]{pluriel2022}. Furthermore, we perform separate retrievals over the first and last part of the transit, allowing us to determine the chemical variation across phase as well as across the terminator. This was possible due to the high precision observations of these two systems available with ESPRESSO, and given that the regions of the atmosphere probed during the transit vary by $\gtrsim30^\circ$ \citep{wardenier2022}. For each half of the terminator we calculate separate phase-dependent limb darkening coefficients. The coefficients vary across the transit because the morning/leading terminator transits the brighter regions of the star during the first half of the transit and thus has a greater contribution, while the second half of the transit has a greater contribution from the evening/trailing side. We maintain $\Delta K_\mathrm{p}$, $\Delta V_\mathrm{sys}$ and $\delta V_\mathrm{wind}$ as shared parameters which remain the same for both halves of the terminator. We perform the HyDRA-2D retrievals including only the species which were detected in the spatially-homogeneous and phase-unresolved HyDRA-H retrievals of these planets in order to keep them computationally efficient. Further details on the setup of the retrievals can be found in \citet{gandhi2022}.

\subsection{Target sample}
We retrieve the atmospheres of six UHJs with publicly available high precision high resolution observations of their terminators: WASP-76~b, MASCARA-4~b, MASCARA-2~b, WASP-121~b, HAT-P-70~b and WASP-189~b. These ultra-hot Jupiters have equilibrium temperatures ranging between $\sim$2210-2640~K (see Table~\ref{tab:planet_params}), where numerous refractory species are expected to be gaseous \citep{lothringer2018}. We used ESPRESSO observations for WASP-76~b, MASCARA-4~b and WASP-121~b, HARPS-N for MASCARA-2~b, and HAT-P-70~b, and a combination of HARPS and HARPS-N for WASP-189~b. These high precision transit observations have previously been used to detect a range of species in the atmospheres of each of the planets, most notably Fe, and are detailed below.

The reduced WASP-76~b data were downloaded from the Data and Analysis Center for Exoplanets (DACE) database\footnote{\url{https://dace.unige.ch/dashboard/index.html}}. These data were originally published in \citet{ehrenreich2020}, which showed clear absorption of Fe but also variability in its signal with orbital phase, potentially due to rain-out of Fe and/or clouds \citep{savel2022, gandhi2022}. These ESPRESSO observations have also been used to detect numerous other atomic and ionic species in the atmosphere \citep{tabernero2021, kesseli2022}. The MASCARA-4~b transit observations were recently published in \citet{zhang2022}, showing numerous atoms and ions. The fully reduced 1D spectra of WASP-121~b from ESPRESSO and WASP-189~b from HARPS were downloaded from the ESO archive\footnote{\url{http://archive.eso.org/scienceportal/home}}, while the HARPS-N spectra were downloaded from the TNG archives\footnote{\url{http://archives.ia2.inaf.it/tng/}}. The ESPRESSO observations of WASP-121~b \citep{borsa2021} added new detections to an already rich spectrum containing many neutral and ionised atoms from many different spectrographs \citep{cabot2020, ben-yami2020, hoeijmakers2020_W121, merritt2021}. The HAT-P-70~b observations were originally published in \citet{bello-arufe2022}, while WASP-189~b observations were published by \citet{prinoth2022} and showed the first conclusive detection of TiO at high spectral resolution. Finally, the HARPS-N observations of MASCARA-2~b have been used in many previous studies \citep{casasayas-barris2018, casasayas-barris2019, stangret2020, nugroho2020}, also showing a large range of atoms and ions. These atomic detections were further confirmed in \citet{hoeijmakers2020} using the EXPRES spectrograph. 

With the higher spectral resolution and signal-to-noise available from ESPRESSO/VLT, we chose to perform spatially-resolved and phase-separated HyDRA-2D retrievals for both WASP-76~b and WASP-121~b. The phase ranges of the retrievals are given in Table~\ref{tab:wind_constraints}. We leave out the phases near 0 to avoid interference with the Doppler shadow of the star, where the planetary signal and the stellar signal have the same velocity component and hence cannot be straightforwardly separated \citep[see e.g.,][]{gandhi2022}.

The observations of MASCARA-4~b are also obtained with the ESPRESSO spectrograph, but this target had strong stellar pulsations as well as Rossiter-Mclaughlin effects and thus we only perform spatially-homogeneous retrievals for this planet. Due to the lower spectral resolution of HARPS-N and HARPS, we also only perform spatially-homogeneous retrievals for the other targets. For MASCARA-2~b and HAT-P-70~b, only upper limits on the planetary masses are available, and hence we include the surface gravity of the planet, log(g), as an additional free parameter in our retrievals for these two planets. This parameter is necessary in order to avoid biases, as the atmospheric scale height is sensitive to the choice of surface gravity and hence planetary mass.

\subsection{Data analysis}\label{sec:data_analysis}

We cleaned and processed all the data following the procedure of \citet{gandhi2022}. We corrected each spectrum for telluric absorption using molecfit \citep{smette2015}, which has been shown to effectively remove contamination from H$_2$O and O$_2$ at visible wavelengths in all but the strongest absorption lines \citep{allart2017}. We masked out any regions of the spectra where transmission through Earth's atmosphere was less than 20\% and hence the molecfit correction returned poor results. We perform the following cleaning steps for each transit observation separately (totalling 13 transits for the six planets). We shifted the spectra to the stellar rest frame using the barycentric velocity during each exposure, the stellar reflex motion (calculated using K$_*$ from Table \ref{tab:planet_params}), and the systemic velocity (V$_\mathrm{sys}$; see Table \ref{tab:planet_params}). We performed a 5-sigma clipping and applying a Gaussian high-pass filter with a full-width half maximum of 100 km s$^{-1}$ to create highly uniform grids of spectra. We then interpolated the spectra onto a single wavelength grid for each night, created by averaging all the wavelengths together during each observation set. Finally, we divided each spectrum by the average out-of-transit spectrum to remove any contribution from the host star. Note that given the stability of fibre-fed spectrographs such as ESPRESSO, we are able to model the tellurics and divide by the out-of-transit spectra. In general, other less-stable spectrographs and/or those which use infrared observations will typically require more complicated filtering, such as Principal Component Analysis.

\begin{figure*}
\centering
	\includegraphics[width=0.49\textwidth,trim={0cm 0cm 0cm 0},clip]{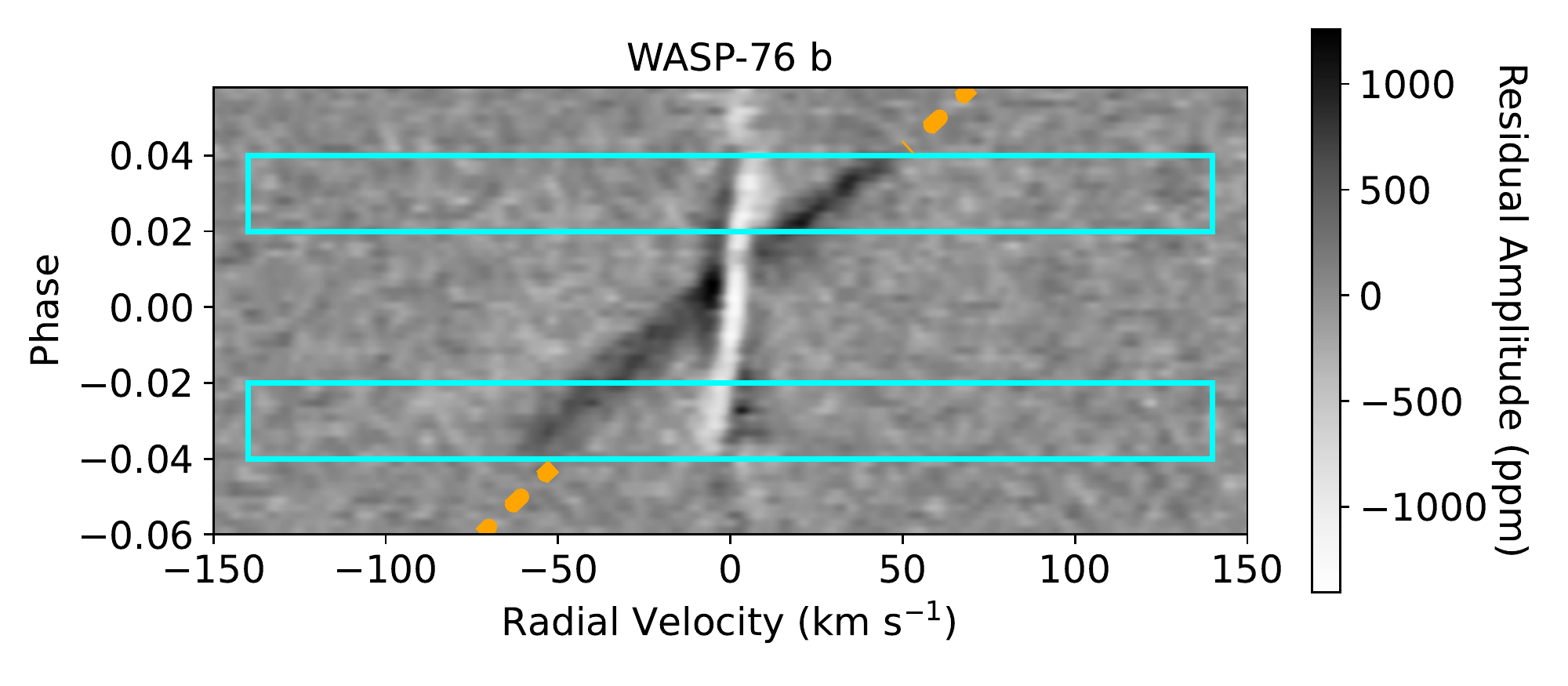}
	\includegraphics[width=0.49\textwidth,trim={0cm 0cm 0cm 0},clip]{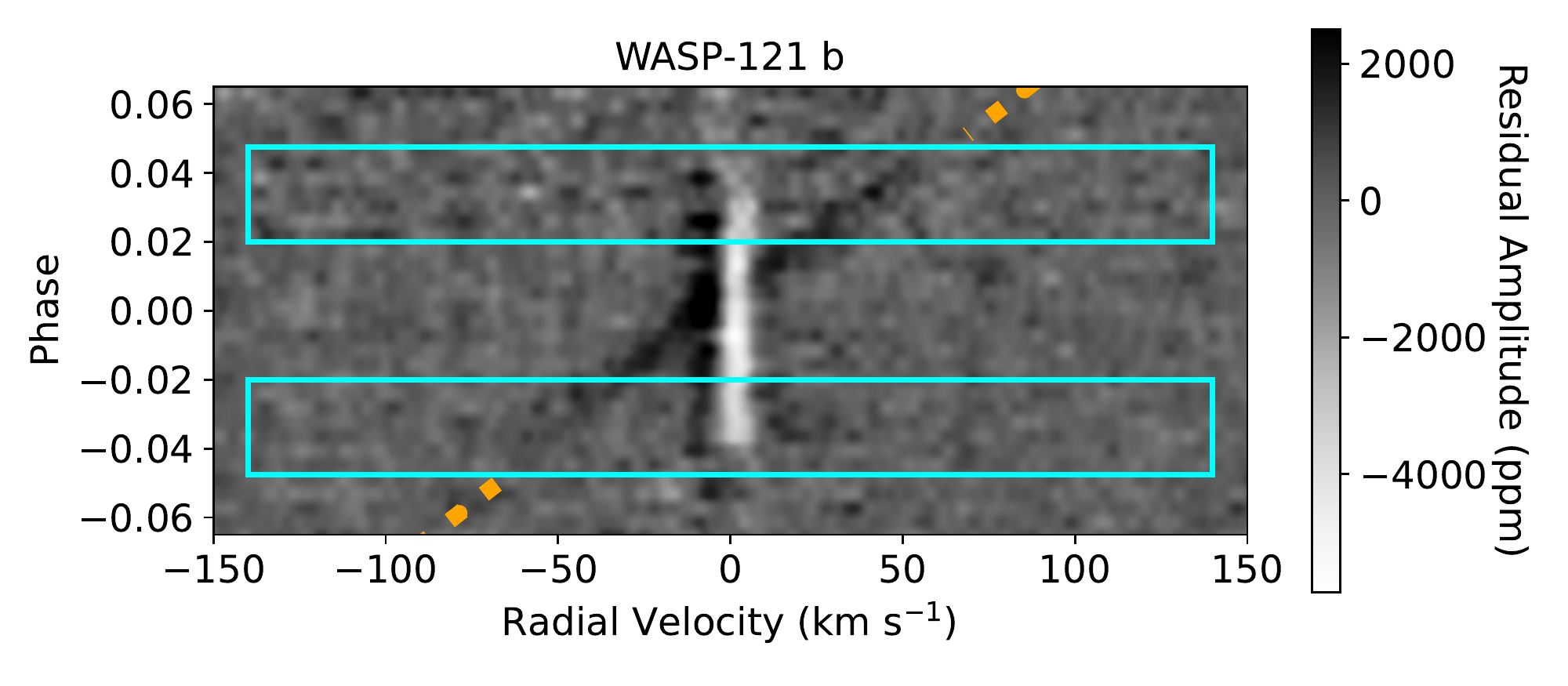}
	\includegraphics[width=0.49\textwidth,trim={0cm 0cm 0cm 0},clip]{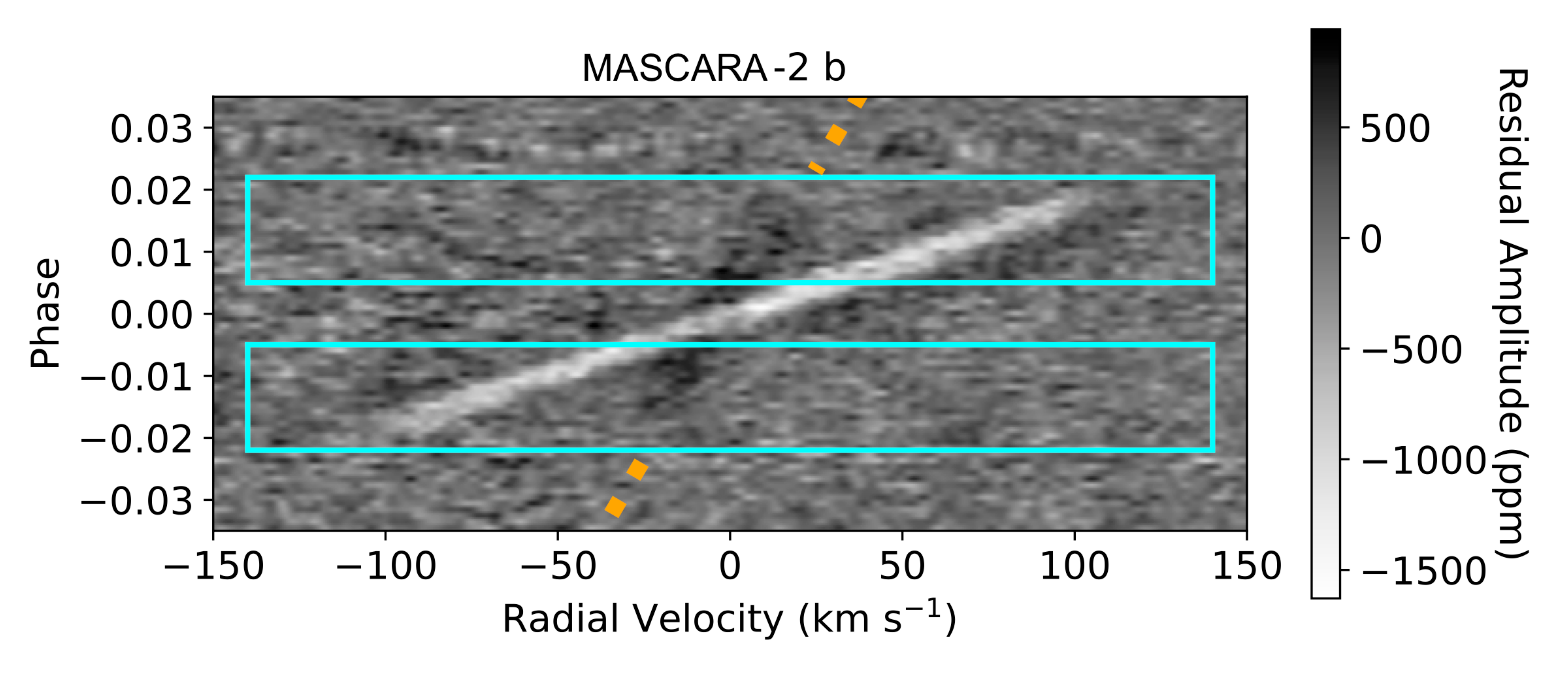}
	\includegraphics[width=0.49\textwidth,trim={0cm 0cm 0cm 0},clip]{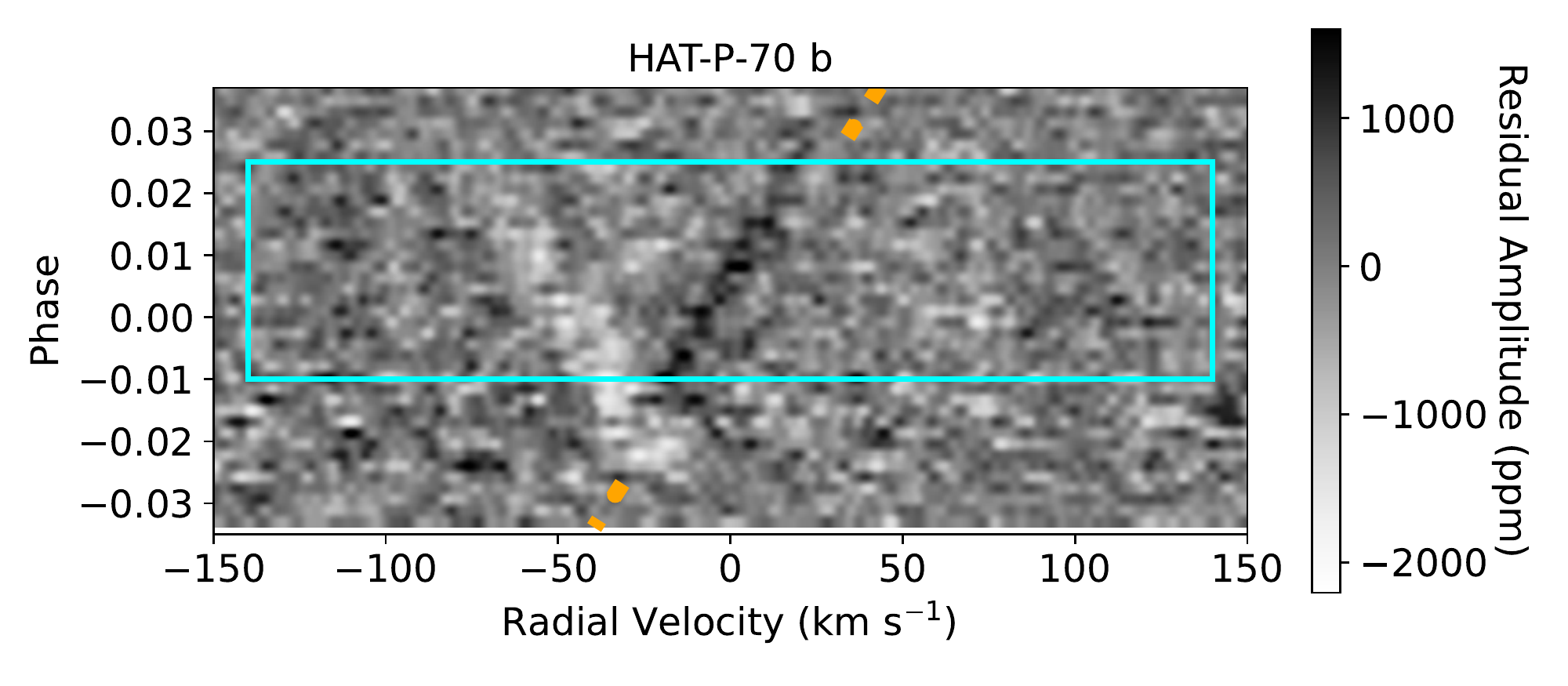}
    \caption{Combined Fe I cross correlation grids for all observed transits for WASP-76~b, MASCARA-2~b, WASP-121~b and HAT-P-70~b. The dotted orange lines indicate the expected velocity of each exoplanet; positive (black) residuals along the orange dotted lines clearly show the detection of Fe I in each exoplanet with a SNR$>$8 in every case. The cyan boxes indicate the regions that we used for the retrievals, chosen to avoid any spectra where the RM and CLV effects directly overlapped with the exoplanet's spectrum.}     
\label{fig:dataCCFs}
\end{figure*}

\begin{figure*}
\centering
	\includegraphics[width=0.49\textwidth,trim={0cm 0cm 0cm 0},clip]{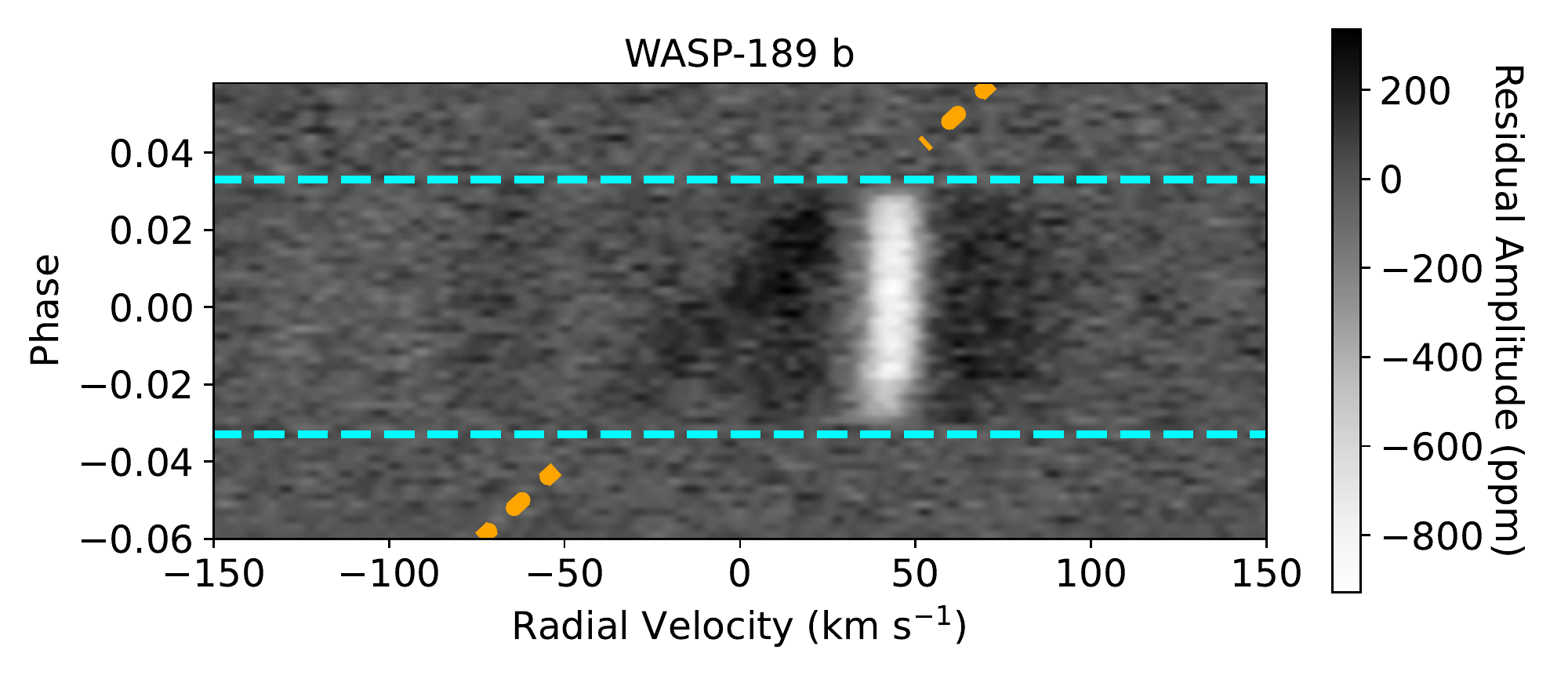}
	\includegraphics[width=0.49\textwidth,trim={0cm 0cm 0cm 0},clip]{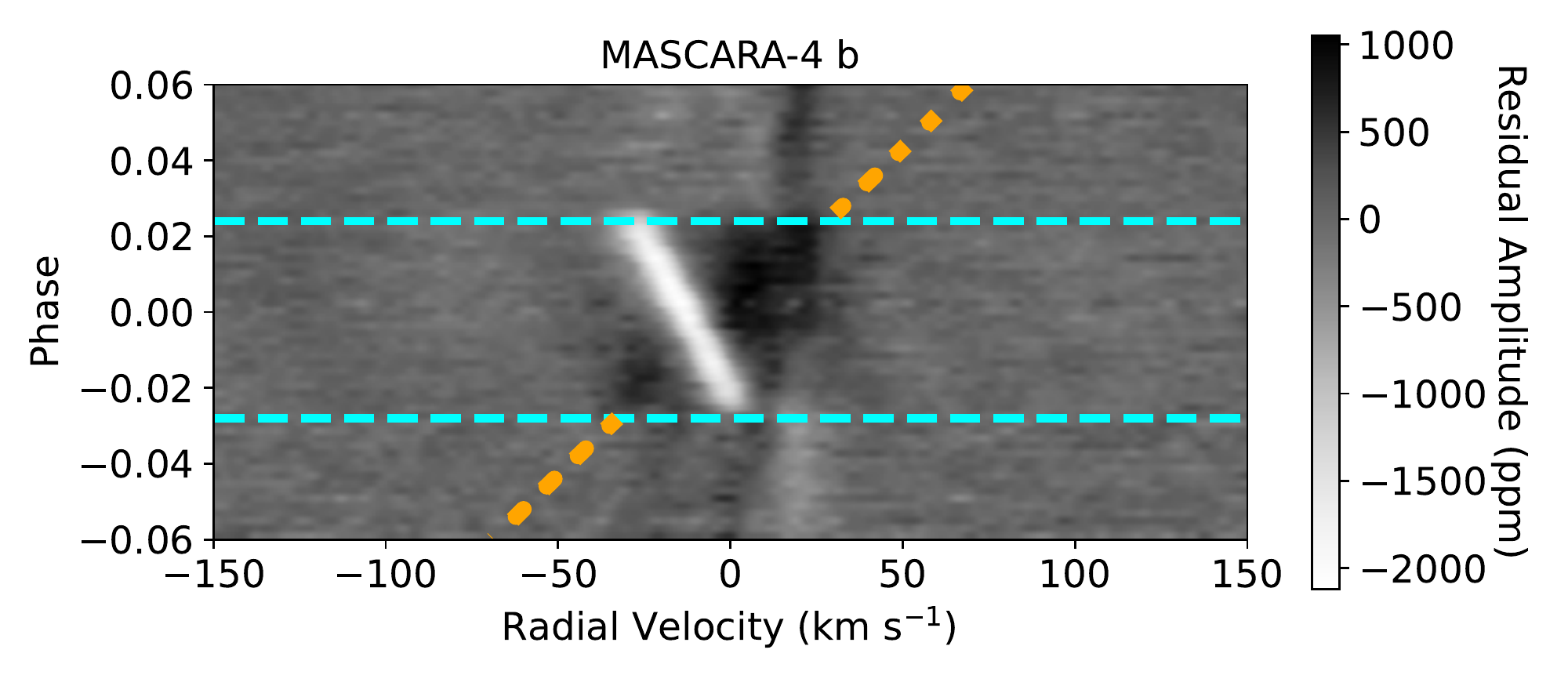}
	\includegraphics[width=0.49\textwidth,trim={0cm 0cm 0cm 0},clip]{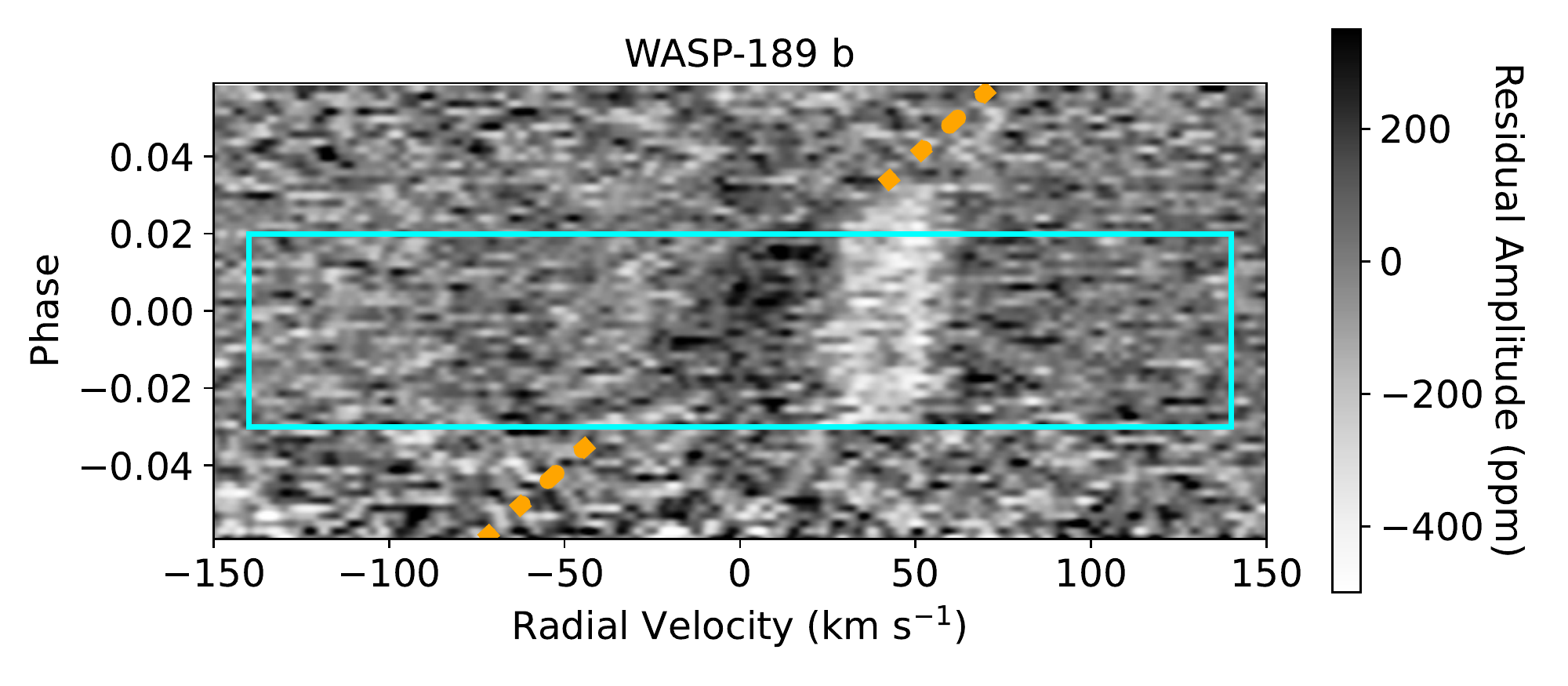}
	\includegraphics[width=0.49\textwidth,trim={0cm 0cm 0cm 0},clip]{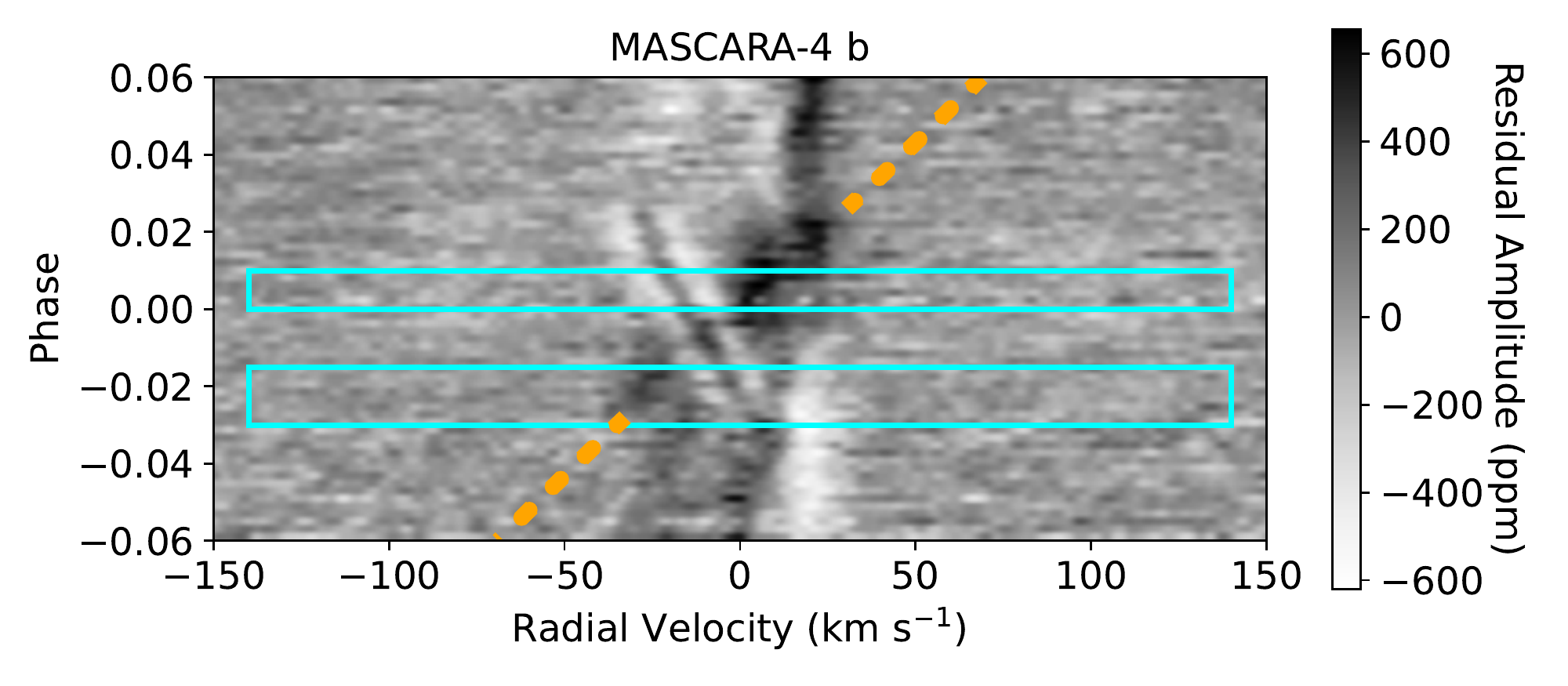}
    \caption{Similar cross correlation functions as Figure \ref{fig:dataCCFs} but for MASCARA-4~b and WASP-189~b, which required a correction for Rossiter-Mclaughlin (RM) effect and centre-to-limb variations (CLV). The top panels show the uncorrected versions, while in the bottom panels our RM and CLV corrections have been applied. In both planets, significant residuals still remain after our corrections, but the amplitude of the stellar contamination is significantly reduced and the planet's signal can be more clearly extracted. Our retrievals using the corrected data converge to the planet's known velocity, ensuring that the signal and molecules constrained are due to the planet. MASCARA-4~b also shows clear contamination due to stellar pulsations (residuals moving from black to white in the vertical direction near 20 km s$^{-1}$). We do not use spectra in the region where the residuals due to stellar pulsations are the most contaminated ($\phi=$ 0.01$-$0.03). }     
\label{fig:dataCCF_RM}
\end{figure*}

The Rossiter-Mclaughlin effect (RM) in combination with centre-to-limb variations (CLV) create residual structure in the cross correlation grid (often called the Doppler shadow), which could influence the results of our retrievals if not removed. Previous methods to correct the Doppler shadow have proven effective in regions where there is not an exact overlap between the shadow and the planet's signal, but struggle to produce results clean enough to detect the planet's signal when there is an exact overlap, especially for cross correlation analysis \citep[e.g.,][]{casasayas-barris2021,casasayas-barris2022}. In addition, the planetary radius appears larger at these regions of overlap, and therefore we chose phase ranges that excluded spectra that had direct overlap with the Doppler shadow. For many of our targets, this contamination was limited to only a few spectra, and the majority of the phases are still included in the retrievals (see Table~\ref{tab:wind_constraints}). Figure \ref{fig:dataCCFs} shows cross correlation grids for each target and an Fe line list, revealing a clear signal in each exoplanet's atmosphere.

We find that for both MASCARA-4~b and WASP-189~b, while the Doppler shadow only directly overlaps with a small portion of the spectra, the rapid stellar rotation creates large wings of positive residuals around the Doppler shadow that bias our retrievals (see Figure \ref{fig:dataCCF_RM}). To correct the RM and CLV effects, we followed \citet{yan2018} and \citet{casasayas-barris2021, casasayas-barris2022} and simulated stellar observations at 10 different limb darkening angles and velocities across the stellar surface using the Spectroscopy Made Easy (SME) package \citep{valenti1996}. To create synthetic observation of the Doppler shadow, we divided the surfaces of the stars in $0.01 \times 0.01 R_*$ pieces and assigned each one a velocity based on the known $v \sin i$s and limb darkening angles. We then combined all the spectra, excluding those that were covered by the planet at the time of each observation, and performed the same cleaning steps on the synthetic observations as we did on the data. Finally, we subtracted these synthetic RM and CLV model spectra from the data to produce the cleaned spectra. The cleaned CCFs are shown in Figure \ref{fig:dataCCF_RM}. As noted in previous works, this method does not remove all contamination from the RM and CLV effects \citep[e.g.,][]{casasayas-barris2022, zhang2022}, but does noticeably improve the contamination. Even using the corrections, the regions where the RM effect directly overlaps with the planet's signal is still highly contaminated and so we still limit the phase ranges used for our retrievals to avoid the direct overlap (see Table~\ref{tab:wind_constraints}). We found that the retrievals performed on the cleaned data converged to the planet's known velocity (K$_p$) for both cases, while retrievals performed on the uncleaned data often converged to K$_p$ values significantly different than the known values due to the RM and CLV residuals. 

In Figure \ref{fig:dataCCF_RM}, MASCARA-4~b also shows residuals due to stellar pulsations, which were found to be similar across the two nights of observations. \citet{zhang2022} assumed that the stellar pulsations remained constant in radial velocity with time, and so removed them by fitting Gaussians to the out-of-transit CCFs. The retrieval set up does not allow for fitting the residuals after cross correlation. We choose to instead omit the phase range that is most contaminated by stellar residuals ($\phi=0.01-0.03$). Even with this limited range of phases, the retrievals quickly converge on the planet's known velocity, and so we are confident of our detections. We do note that the width of the CCFs (parameterized by $\delta$V$_{wind}$) is likely biased due to the pulsations. Furthermore, as the stellar pulsations affect all species, we expect the ratio of abundances (i.e. Mg/Fe, etc.) to be more reliable than absolute abundances for the case of MASCARA-4~b. 

The cleaned spectra in the specified phase ranges were compared to the models in our retrieval framework. Following \citet{gandhi2022}, we convolved the models to the instrumental resolution and shifted them in velocity at each phase given their $\Delta K_\mathrm{p}$ and $\Delta V_\mathrm{sys}$, which are free parameters in the retrieval. The model spectra are also scaled for limb darkening at each phase using the quadratic limb darkening coefficients for each system. Finally, the spectra are rotationally broadened and a high pass filter is applied to the model to reproduce the steps in the data analysis. 

The processed model and cleaned spectra are compared using a cross-correlation to log-likelihood map \citep{brogi2019}. The log-likelihood value for a given model compared against the data is computed with
\begin{equation}
    \log L = - \frac{N}{2} \log(s_f^2 + s_g^2 - 2R),
\end{equation}
where
\begin{equation}
    s_f^2 = \frac{1}{N}\sum_{n=0}^N f^2(n),
\end{equation}
\begin{equation}
    s_g^2 = \frac{1}{N}\sum_{n=0}^N g^2(n-s),
\end{equation}
\begin{equation}
    R = \frac{1}{N} \sum_{n=0}^N f(n)g(n-s).
\end{equation}
Here, $f$ and $g$ are the data and model respectively for a model with $N$ spectral points, and $s$ represents the wavelength offset. The overall value of the log-likelihood is the sum over all orders and phases. For planets where multiple transits were observed the log-likelihoods of each transit were added to obtain the overall value. Our Bayesian analysis is performed with the Nested Sampling algorithm MultiNest \citep{feroz2008, feroz2009, buchner2014}. 

\begin{figure*}
\centering
	\includegraphics[width=\textwidth,trim={0cm 0cm 0cm 0},clip]{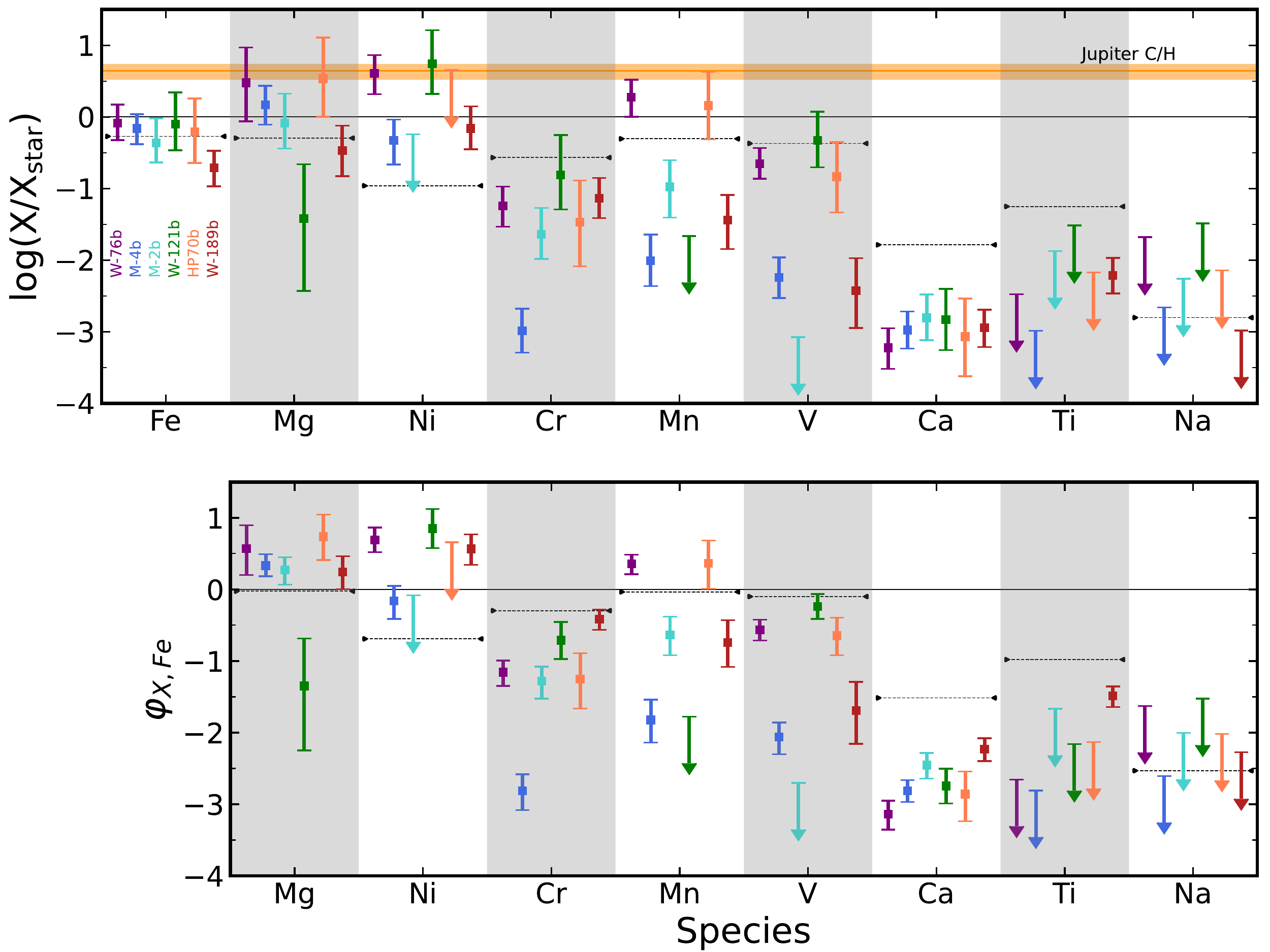}
    \caption{Top panel: Volume mixing ratios and 1$\sigma$ errors of each species relative to the stellar abundance for the planets in our survey, ordered by planetary equilibrium temperature. Bottom Panel: Values of $\varphi_\mathrm{X,Fe}$ and 1$\sigma$ errors for each species, defined as $\varphi_\mathrm{X,Fe} = \log(\mathrm{X}/\mathrm{Fe}) - \log(\mathrm{X_{\odot}}/\mathrm{Fe_{\odot}})$, where X and Fe refer to the each species' volume mixing ratios and $\mathrm{X_{\odot}}$ and $\mathrm{Fe_{\odot}}$ refer to their solar abundance (see section~\ref{sec:relative_abundances}). Where no significant constraints were retrieved, we show the 2$\sigma$ upper limit. Note that the Ti value is the sum of the constraints for atomic Ti, and molecular TiO and TiH. The dashed lines for each species indicate the expected abundance in chemical equilibrium at 3000~K and 0.1~mbar pressure (see Figure~\ref{fig:eqm_chem}).}     
\label{fig:chem}
\end{figure*}

\begin{table*}[]
   \centering
    \def\arraystretch{1.5}
\begin{tabular}{c|cccccc}
Species & \multicolumn{1}{c}{\bf{WASP-76~b}} & \multicolumn{1}{c}{\bf{MASCARA-4~b}} & \multicolumn{1}{c}{\bf{MASCARA-2~b}} & \multicolumn{1}{c}{\bf{WASP-121~b}} & \multicolumn{1}{c}{\bf{HAT-P-70~b}} & \multicolumn{1}{c}{\bf{WASP-189~b}}\\
\hline
Fe & $-4.24^{+0.26}_{-0.24}$ & $-4.68^{+0.20}_{-0.22}$ & $-4.90^{+0.34}_{-0.27}$ & $-4.49^{+0.44}_{-0.36}$ & $-4.79^{+0.46}_{-0.44}$ & $-4.94^{+0.24}_{-0.26}$\\
Mg & $-3.65^{+0.49}_{-0.54}$ & $-4.32^{+0.27}_{-0.28}$ & $-4.60^{+0.41}_{-0.36}$ & $-5.78^{+0.76}_{-1.01}$ & $-4.02^{+0.57}_{-0.53}$ & $-4.67^{+0.35}_{-0.36}$\\
Ni & $-4.81^{+0.26}_{-0.29}$ & $-6.11^{+0.29}_{-0.34}$ & $<-6.04$ & $-4.91^{+0.47}_{-0.42}$ & $<-5.18$ & $-5.65^{+0.31}_{-0.29}$\\
Cr & $-7.24^{+0.27}_{-0.29}$ & $-9.35^{+0.31}_{-0.31}$ & $-8.02^{+0.37}_{-0.34}$ & $-7.04^{+0.56}_{-0.48}$ & $-7.89^{+0.58}_{-0.62}$ & $-7.21^{+0.28}_{-0.28}$\\
Mn & $-5.93^{+0.25}_{-0.27}$ & $-8.58^{+0.37}_{-0.35}$ & $-7.57^{+0.37}_{-0.43}$ & $<-8.10$ & $-6.47^{+0.47}_{-0.47}$ & $-7.72^{+0.35}_{-0.40}$\\
V & $-8.36^{+0.22}_{-0.21}$ & $-10.31^{+0.28}_{-0.29}$ & $<-11.17$ & $-8.27^{+0.40}_{-0.37}$ & $-8.96^{+0.48}_{-0.50}$ & $-10.20^{+0.45}_{-0.52}$\\
Ca & $-8.52^{+0.27}_{-0.29}$ & $-8.64^{+0.26}_{-0.26}$ & $-8.49^{+0.33}_{-0.31}$ & $-8.36^{+0.43}_{-0.42}$ & $-8.79^{+0.53}_{-0.55}$ & $-8.31^{+0.25}_{-0.27}$\\
$^*$Ti & $<-9.16$ & $<-10.04$ & $<-8.95$ & $<-8.43$ & $<-9.28$ & $-8.97^{+0.24}_{-0.25}$\\
Na & $<-7.07$ & $<-8.42$ & $<-8.04$ & $<-7.12$ & $<-7.96$ & $<-8.45$\\
\end{tabular}
    \caption{Elemental atmospheric abundance constraints (given in log(volume mixing ratio)) and their 1$\sigma$ errors for the planets in our survey. Where no significant constraints were retrieved, we show the 2$\sigma$ upper limit. $^*$Ti value is the sum of the constraints for atomic Ti, and molecular TiO and TiH.}
    \label{tab:chem_constraints}
\end{table*}

\section{Results and discussion}\label{sec:results}

In this section we discuss the results from our retrieved abundances and wind profiles for each planet and compare their constraints. We firstly explore the chemical constraints for the eleven species for each of the six planets in our survey, followed by a discussion of relative abundance constraints with high-resolution spectroscopy. We then discuss the HyDRA-2D retrievals performed on WASP-76~b and WASP-121~b, and the abundance variation of each species in the different regions of the terminator. Finally, we compare and contrast the wind profiles we obtain across the sample and discuss the mass constraints for MASCARA-2~b and HAT-P-70~b.

\begin{figure*}
\centering
	\includegraphics[width=\textwidth,trim={0cm 0cm 0cm 0},clip]{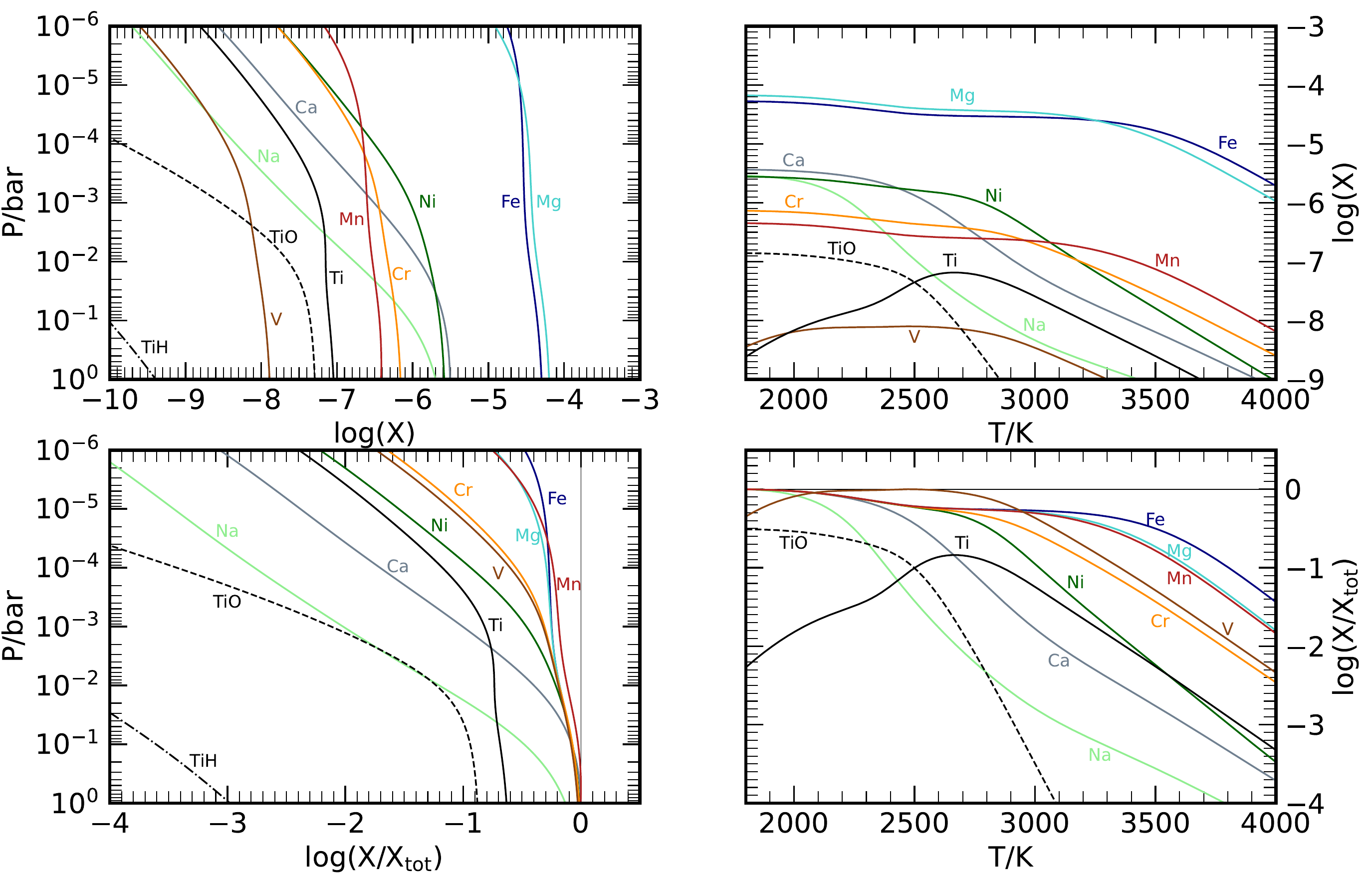}
    \caption{Chemical equilibrium abundances (in log(volume mixing ratio)) of each species at solar elemental abundance \citep{asplund2009}. The top left panel shows the variation of the volume mixing ratio of each species with the pressure at a fixed isothermal temperature of 3000~K. The top right panel shows the variation of volume mixing ratio with the temperature at a fixed isobaric pressure of 0.1~mbar. The bottom panels for each side show the corresponding fraction of each species to the total atom content in the atmosphere, $\mathrm{X_{tot}}$.}    
\label{fig:eqm_chem}
\end{figure*}

\subsection{Chemical abundances}\label{sec:abundances}

We performed spatially homogeneous retrievals for the six planets in our survey, WASP-76~b, MASCARA-4~b, MASCARA-2~b, WASP-121~b, HAT-P-70~b and WASP-189~b. We retrieve the volume mixing ratios of eleven species, Fe, Mg, Ni, Cr, Mn, V, Ca, Ti, TiO, TiH and Na. These species were chosen as they are expected to be the most prominent in the optical and have often been detected in the atmospheres of UHJs \citep[e.g.,][]{merritt2021, kesseli2022}. Figure~\ref{fig:chem} and Table~\ref{tab:chem_constraints} show the abundance constraints for each species in each of the planets, with the overall Ti value the sum of the constraints from atomic Ti, TiO and TiH. We note that the abundances for each species correspond to the photospheric constraints at the terminator of these planets. The majority of species show clear abundance constraints, with 2$\sigma$ upper limits for the species which did not show any peak in the posterior distribution. From these, we can see that Fe is the most clearly constrained species, in line with expectations given its high abundance and opacity. Planets such as WASP-76~b and MASCARA-4~b show the tightest constraints given their multiple nights of high signal-to-noise observations.

To compare our retrieved abundances we modelled the chemical equilibrium abundances of the species in our survey with \texttt{FastChem} \citep{Stock2018}. This chemical equilibrium code assumes that the chemical reaction rates are directly proportional to the concentrations of all reactants, also known as the \textit{law of mass action}. By minimising the Gibbs free energy of the closed atmosphere system the abundances of various molecules, atoms, and ions are calculated at a given pressure and temperature in gas-phase. These models account for ionisation of species as well as the formation and interaction of molecules in the atmosphere, and the chemical element composition assumes solar abundances from \cite{asplund2009}. The abundances for each species are shown in Figure~\ref{fig:eqm_chem} as a function of pressure and temperature. We show the expected volume mixing ratio as well as the ratio of each of species to the total atom content, $\mathrm{X_{tot}}$, of the atmosphere. Typically, high-resolution ground-based observations in the optical probe pressures near $\sim$0.1~mbar \citep[e.g.,][]{maguire2023}.

The chemical equilibrium models show that as the temperature increases or the pressure decreases, the abundance of each species decreases due to ionisation. Hence we expect lower abundances of the neutral species as the temperature increases or as we probe higher up in the atmosphere. For species such as Fe and Mg the overall ionisation is low until we get to very high temperatures ($\gtrsim$3500~K), but for others such as Ca, Ti and Na the abundance drops significantly at temperatures $\gtrsim$2700~K and with decreasing pressure. For Ti, the dominant carrier of the atom is expected to be molecular TiO up to $\sim2500$~K, after which point atomic Ti becomes stronger as the TiO dissociates, before that too depletes at higher temperatures as the neutral Ti ionises.

\begin{figure}
\centering
	\includegraphics[width=0.49\textwidth,trim={0.0cm 0cm 0cm 0},clip]{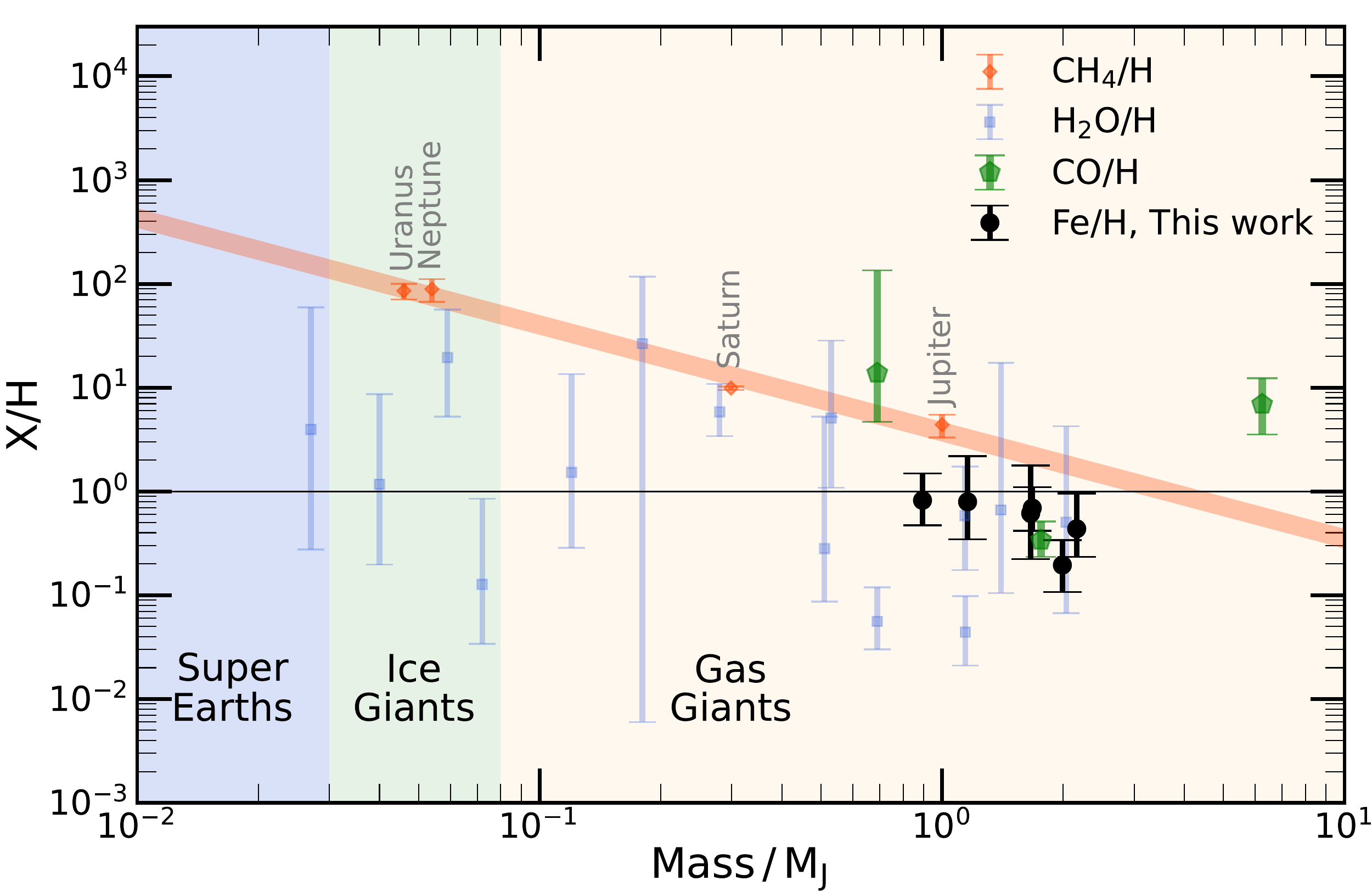}
    \caption{Mass-metallicity relation for known exoplanets and the Solar System, corrected for the stellar metallicity of each system. The Fe constraint for each planet from this work is shown in black. The orange region indicates the Solar System trend. We also show the H$_2$O/H (blue) and CO/H (green) constraints for a range of exoplanets obtained with low-resolution and high-resolution observations respectively \citep{welbanks2019, guillot2022}.}
\label{fig:mass_met}
\end{figure}

\subsubsection{Fe}

We compared the Fe abundance of each planet relative to their host star's metallicity, shown in Table~\ref{tab:planet_params}. The constrained abundances of Fe for all of the planets are broadly consistent with the stellar value. This may indicate that these planets may have formed under similar conditions. Only WASP-189~b showed an Fe constraint with the 1$\sigma$ error below the stellar Fe/H value, by $\sim$0.7~dex. The chemical equilibrium models in Figure~\ref{fig:eqm_chem} show that the majority of the iron remains as neutral Fe for typical photospheric conditions unless the temperature becomes very high ($\gtrsim$3500~K). As WASP-189~b is the hottest planet in our sample (see Figure~\ref{fig:PT}), the lower value of Fe may be due to a significant amount of Fe being ionised. Alternatively, the lower Fe abundance could hint at a mass-metallicity trend as WASP-189~b is the most massive planet in our sample that has a known mass determination (see Table~\ref{tab:planet_params}). 

Figure~\ref{fig:mass_met} shows the mass-metallicity relation from the Fe/H for each of the planets in our survey. We determine the masses of MASCARA-2~b and HAT-P-70~b from the median value of log(g) retrieved for each (see section~\ref{sec:masses}). We see a weak trend of decreasing abundance with mass in line with the Solar System trend in C/H, but significant trends are difficult to ascertain given the limited mass range of these planets. Our findings are consistent with recent work from low-resolution retrievals, which showed at or near stellar abundances of alkali metals in hot Jupiters \citep[e.g.,][]{welbanks2019}. On the other hand, recent high-resolution dayside observations of MASCARA-2~b and other ultra-hot Jupiters have shown evidence for super-stellar abundances \citep{cont2022, kasper2022}, indicating that there may be some variation in abundances of Fe in UHJs. Neutral Fe remains gaseous down to temperatures of $\sim$1350~K at 0.1~mbar (see Table~\ref{tab:cond}), does not strongly ionise until high temperatures (see Figure~\ref{fig:eqm_chem}), and is the dominant carrier of iron for these UHJs \citep{visscher2010}. Therefore, neutral Fe is likely to be a good measure of the bulk Fe composition of such targets. However, this also holds true for other species such as neutral Mg, Cr or Mn, which are also relatively stable and the dominant carrier of the atom. Hence using their abundance as a proxy for atmospheric metallicity for the planets in our survey could potentially result in differing conclusions (see below), and highlights the difficulty of using the abundance of a single species to probe bulk metallicity.

The atmospheric abundance of refractory species such as Fe is a particularly important quantity as it is an indicator of the rock composition accreted during the planet's formation \citep{lothringer2021}. \citet{chachan2022} further demonstrate how measuring the abundances of refractory species, specifically Fe, can break degeneracies that exist when trying to determine the formation location of an exoplanet when using only volatile species, such as C and O rich species. In the future, our results of measured refractory abundances can be combined with abundances of volatiles from near infrared observations, to constrain how the population of ultra-hot Jupiters formed and evolved, as was recently done with dayside spectra \citep{kasper2022}. 

\subsubsection{Mg, Ni, Cr and Mn}

The abundances of Mg, Ni, Cr and Mn show weaker constraints and greater variation than Fe. The Mg constraints have a wider uncertainty given its relatively few spectral lines in the range of our observations, but these agree well with the results for Fe. Our Ni abundance across the sample shows a trend of super-stellar values, but Cr indicates more sub-stellar abundances, particularly for MASCARA-4~b. In some of our retrievals we constrain very low abundances (log(Volume Mixing Ratio) $<$ -9) for Cr owing to their strong optical opacity and the high-precision observations. \citet{pelletier2023} find a similar trend in their retrievals of WASP-76~b and suggest that increased Ni abundance could be due to an impact during planet formation. However, neither constraints for Ni or Cr are as significant as Fe given the generally weaker absorption of these species making precise abundance constraints and hence interpretations more difficult. Ni has a higher ionisation fraction than Fe, Mg, Cr or Mn (see Figure~\ref{fig:eqm_chem}), and its higher retrieved abundances may also be an indicator of vertical mixing driving more neutral Ni towards the photosphere at higher altitudes. The Mn abundance also shows variation across our sample, with WASP-76~b and HAT-P-70~b indicating super-stellar abundances but with the other planets indicating sub-stellar values. Hence further observations of the terminator of UHJs is required to explore these variations in abundances further. 

We note that we are able to constrain Ni and Mn in the atmosphere of MASCARA-4~b, but \citet{zhang2022} were not able to clearly pick this up in cross-correlation. The retrieval is able to explore a wide range of chemical abundances, temperature profiles and opacity decks, which increases the parameter space probed which may result in better constraints. This is particularly important for species such as Mn, which was constrained to be $\sim$2 orders of magnitude lower than its expected value. Furthermore, our retrieval fits for all species simultaneously, meaning that masking of spectral features from Ni and Mn by other stronger species (e.g. Fe) is accounted for. The observations of MASCARA-4 had strong RM+CLV which we did model, and we ensured that we used phases which avoided the stellar residuals, but we caution that our constraint for this planet may be affected by some uncorrected stellar residuals. \citet{zhang2022} were also able to detect Na, but our retrievals showed only a weak peak with no clear constraint (see section~\ref{sec:na}), indicating that these two methodologies may have slight advantages and disadvantages for weak chemical signatures depending on the species being probed.

\subsubsection{V}

The V abundance for the planets in our survey also shows more variation, with WASP-76~b, WASP-121~b and HAT-P-70~b showing abundances consistent with the stellar values, but MASCARA-4~b, MASCARA-2~b and WASP-189~b showing significantly lower values or upper limits. This may potentially be due to the formation of VO which depletes the atmosphere of neutral V. We do not include VO as part of our analyses given the lack of reliable high resolution line lists available \citep{deregt2022}. VO possesses a strong optical opacity and is also capable of causing thermal inversions in ultra-hot Jupiters \citep[e.g.,][]{hubeny2003, fortney2008, gandhi2019_inv}. On the other hand, the three planets with the lowest V abundances also possess the strongest RM and CLV, which may be interfering with the V signal in the observations. \citet{cont2022} suggest depleted V abundances for the UHJ WASP-33~b in emission compared with other species, showing that there may be some variation in inferred V abundances in UHJs.

\subsubsection{Ca and Ti}\label{sec:ca_ti}

The chemical equilibrium models show that the neutral Ca and Ti are expected to be reduced by $\sim$1-2 orders of magnitude at a temperature of 3000~K and 0.1~mbar pressure (dashed line in Figure~\ref{fig:chem}). This brings us to a much closer agreement to the retrieved abundances for these species. However, there are still some differences, with Ca and Ti constrained to be even lower than what the chemical equilibrium models predict. This may indicate that the temperature is higher, as this will act to further ionise the Ca and Ti and reduce the retrieved abundances. The temperature profile constraints do indicate that the median temperatures are higher than 3000~K for many of these planets (see Figure~\ref{fig:PT}), but our retrieved abundances do not show a strong dependence with the retrieved temperature profiles. In fact, the Ca abundance in WASP-76~b, the coolest planet in our survey, is the lowest of any of the planets when we account for the host star metallicity (see Figure~\ref{fig:chem}).

A potential explanation for the low abundances of gaseous Ca and Ti bearing species is the formation of CaTiO$_3$ clouds on the night side, depleting the Ca and Ti in the atmosphere due to rain-out. These clouds form at temperatures of $\sim1600$~K at 0.1~mbar (see Table~\ref{tab:cond}), and the night sides of UHJs have been shown to be cool enough for such clouds to precipitate \citep{mikal-evans2022}. As CaTiO$_3$ has the hottest condensation temperatures for the species used in our work, Ca and Ti are the most likely to be depleted due to rain-out on the night side. Hence Ti-bearing species may be sequestered deep in the interior of the planet through cold-trapping \citep[e.g.,][]{spiegel2009}. Recent work has also shown no evidence for Ti or TiO in WASP-121~b \citep{hoeijmakers2022, maguire2023}, or TiO in MASCARA-2~b \citep{johnson2022}.

Across our sample we are only able to constrain Ti-bearing species (both Ti and TiO) in WASP-189~b, as found by \citet{prinoth2022}. This is the hottest exoplanet in our sample with temperature constraints $\sim$3900~K (see Figure~\ref{fig:PT}), and hence least susceptible to condensation of Ti-rich species. The lower equilibrium abundances of neutral species such as Fe at these high temperatures may also reduce the shielding of Ti and TiO spectral features in addition to any effect clouds/quenching may have. TiO is of particular interest given that it has strong optical opacity and can cause thermal inversions in the photosphere \citep[e.g.,][]{hubeny2003, fortney2008, piette2020}.

\begin{figure*}
\centering
	\includegraphics[width=\textwidth,trim={0cm 0cm 0cm 0},clip]{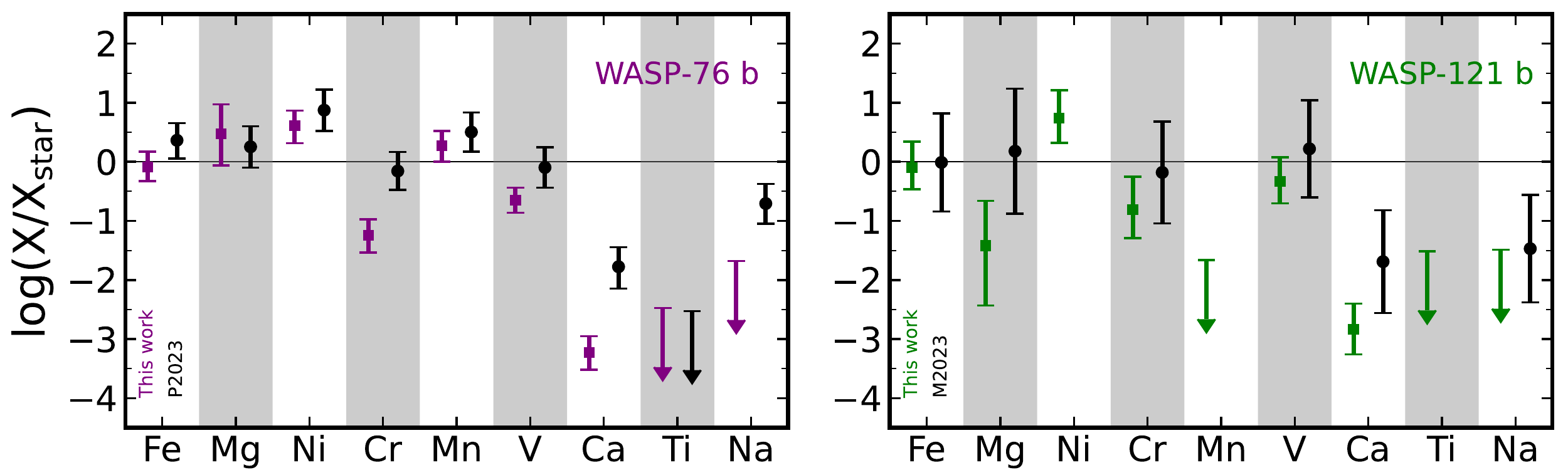}
    \caption{Comparison of our retrieved abundances for each species with \citet{pelletier2023} for WASP-76~b in the left panel and \citet{maguire2023} for WASP-121~b in the right panel.}
\label{fig:compare}
\end{figure*}

Another possibility is vertical chemical gradients in the atmosphere which artificially reduce the constrained abundance. The chemical equilibrium models show that Ca, Na and Ti are expected to have some of the strongest decreases in their abundance with atmospheric height, but our retrieval is not able to capture this given that we assume vertically homogeneous chemistry. Hence the cores of spectral lines, which are generated at lower pressures, will be weaker than expected given the strong ionisation of neutral species in the upper atmosphere, and will therefore drive the retrieved abundances to lower values. In addition, other species containing Ca may form in the atmosphere and deplete the atmosphere of neutral gaseous atoms. Molecules such as CaO have also been shown to be capable of causing stratospheres in the atmospheres of UHJs \citep{gandhi2019_inv}, but are likely to make up a small fraction of the overall Ca budget in equilibrium. Determination of the Ca+ abundance will provide a key indicator to the overall ionisation of neutral Ca and the total Ca content of the atmosphere, as many of these planets have shown very strong Ca+ absorption \citep{merritt2021, kesseli2022, zhang2022, pelletier2023}. However, including this into retrievals may be challenging as they probe the exosphere \citep{maguire2023}.

\subsubsection{Na}\label{sec:na}

Neutral Na is expected to be substantially ionised at the conditions typical for ultra-hot Jupiter photospheres, with the equilibrium models indicating depletion by almost three orders of magnitude at 3000~K and 0.1~mbar (see Figure~\ref{fig:eqm_chem}). Given this lower abundance and the relatively few prominent spectral lines, our retrievals are unable to conclusively constrain Na for any of the planets, but the upper limits on the retrieved abundances are consistent with the equilibrium model predictions, with 2$\sigma$ upper limits around $\log(\mathrm{Na}) \sim -8$ (see Table~\ref{tab:chem_constraints}). We did obtain tentative peaks for Na in WASP-121~b and MASCARA-4~b, hinting at a detection, but in both cases there was a long tail in the abundance distribution that extended to the edge of our prior, and so we conservatively do not report these as confident detections. Other works have been able to constrain Na through other observations (see section~\ref{sec:other_works}). However, given its relatively few spectral lines, direct detection of the spectral features from the Na $\sim$0.589~$\mu$m D lines have proved the most effective to detect Na in these planets \citep[e.g.,][]{seidel2019, casasayas-barris2019, borsa2021, langeveld2022, zhang2022}. Furthermore, the Na feature had a velocity shift relative to other species such as Fe for WASP-76~b \citep{kesseli2022}, as well as a significant velocity offset from the known orbital solution for WASP-121~b \citep{seidel2023}. This may also reduce our constraints given that the retrieval assumes no velocity shifts between species. We note that the strong opacity of the D lines may also result in the absorption features occurring in the non-hydrostatic regions of the atmospheres at very low pressures \citep{hoeijmakers2020_W121, gebek2020}, thereby making accurate inferences of Na abundances more difficult.

\subsection{Comparison with previous works}\label{sec:other_works}

We compared our results to previous retrievals of the terminator of WASP-76~b and WASP-121~b. These used MAROON-X/Gemini-North observations by \citet{pelletier2023} for WASP-76~b and ESPRESSO/VLT observations by \citet{maguire2023} for WASP-121~b. The comparison of the retrieved volume mixing ratios are shown in Figure~\ref{fig:compare}. We find excellent agreement with both studies, but \citet{pelletier2023} did constrain higher abundances for Cr and Ca in WASP-76~b compared with our values. Their work constrained both neutral Ca and ionised Ca+ and showed that the Ca+ dominates over the neutral Ca, indicating that the calcium in the atmosphere is almost completely ionised. They were also able to constrain Na, with an abundance slightly above our upper bound. For WASP-121~b, \citet{maguire2023} retrieved upper limits to Ti similar to ours, but did constrain Na at abundances similar to but slightly above our upper bounds. Furthermore, \citet{gibson2022} constrained Fe, Mg, Cr and V in WASP-121~b with UVES/VLT observations, which are also in good agreement with our abundance constraints. We note that for these studies and our retrievals the relative abundances are more robust (see below). However, given the differences in datasets, analyses and atmospheric modelling, the consistency of these results is encouraging, and a verification of the statistical and modelling frameworks used for high-resolution retrievals.

\begin{figure*}
\centering
	\includegraphics[width=\textwidth,trim={0cm 0cm 0cm 0},clip]{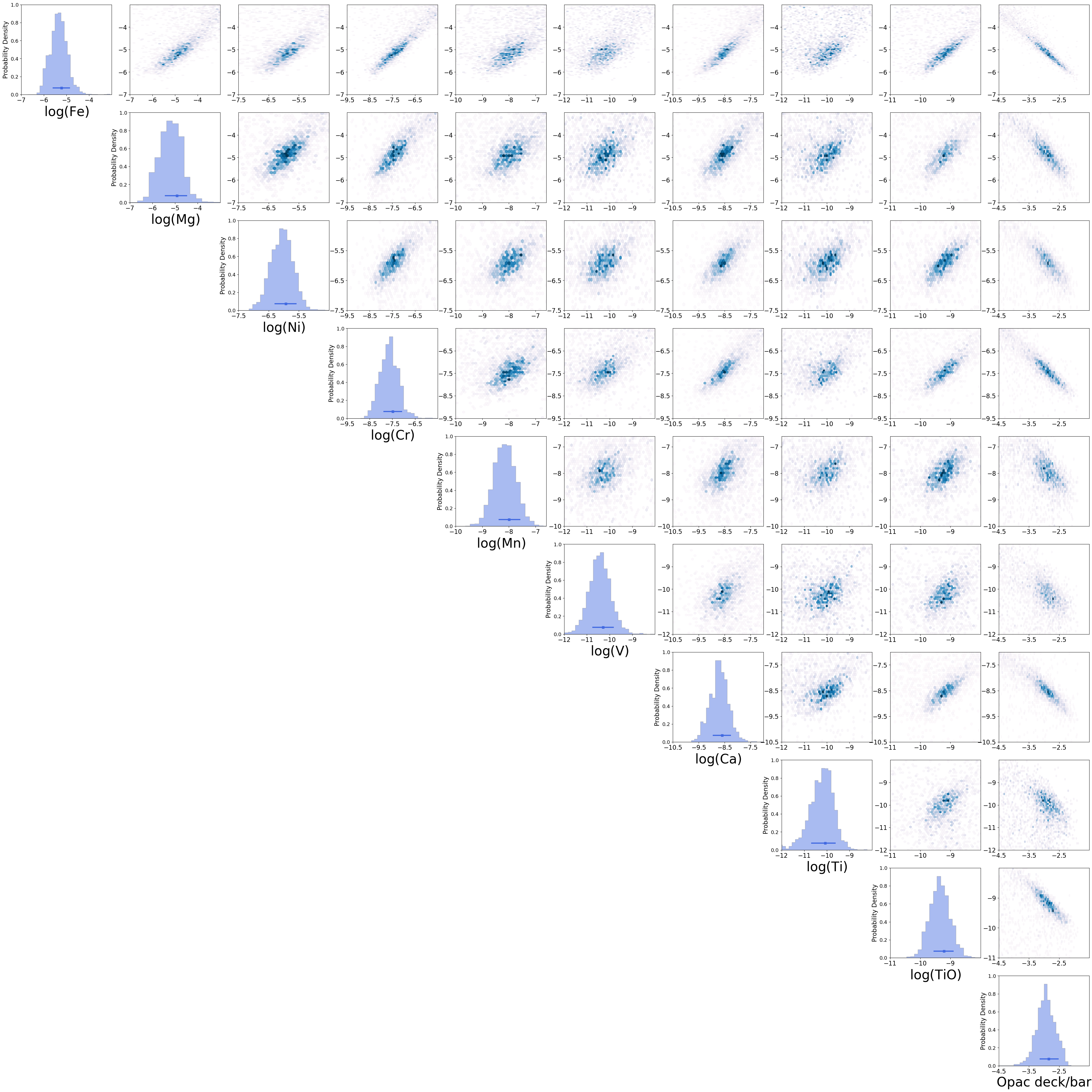}
    \caption{Posterior distribution of the volume mixing ratios of Fe, Mg, Ni, Cr, Mn, V, Ca, Ti, TiO and the opacity deck for the HyDRA-H retrieval of WASP-189~b performed over the $\phi  = -0.033 \,\text{-}\, +0.02$ range. These highlight the dependence of each parameter on the other (see section~\ref{sec:relative_abundances}).}     
\label{fig:corner}
\end{figure*}

\subsection{Constraining relative abundances with high-resolution spectroscopy}\label{sec:relative_abundances}

Figure~\ref{fig:corner} shows the posterior distribution for WASP-189~b, where we clearly constrain the abundances for nine atomic and molecular species and the pressure level of the opacity deck. The posterior shows that the abundance constraints are negatively correlated with the location of the opacity deck, as found previously for Fe for WASP-76~b \citep{gandhi2022}. This is because HRS is more sensitive to the relative strengths of the spectral lines over the continuum (i.e. the opacity deck), rather than the absolute level. Species with strong opacity, such as Fe, TiO and Cr were the most strongly correlated with the opacity deck given their significant number of spectral lines. Therefore, this leads to abundances between species being positively correlated with each other. For some species such as Cr the trend of increasing abundances with increasing Fe is particularly strong (see Table~\ref{tab:chem_fe_ratio_constraints}). This indicates that relative abundance constraints with high resolution spectroscopy are much more precise and robust, as found by previous works \citep[e.g.,][]{gibson2022, maguire2023, pelletier2023}. Such relative abundances over the continuum are also common in low-resolution spectroscopy \citep[e.g.,][]{welbanks2019_degen}, indicating this is a general feature of transmission spectroscopy. 

It is instructive to define a new variable which allows us to clearly explore the relative abundances between species. We define $\varphi_\mathrm{A,B}$ as
\begin{equation}
    \varphi_\mathrm{A,B} \equiv \log(\mathrm{X_A}/\mathrm{X_B}) - \log(\mathrm{X_{star,A}}/\mathrm{X_{star,B}}),
\end{equation}
where $\mathrm{X_A}$ and $\mathrm{X_B}$ are the atmospheric volume mixing ratios, and $\mathrm{X_{star,A}}$ and $\mathrm{X_{star,B}}$ are the stellar volume mixing ratios of species A and B respectively. This allows us determine the relative enhancement of one species over another with respect to the stellar values. Note that this is similar to definition of the stellar metallicity, 
\begin{equation}
    [\mathrm{Fe/H}] \equiv \log(\mathrm{X_{star,Fe}}/\mathrm{X_{star,H}}) - \log(\mathrm{X_{\odot,Fe}}/\mathrm{X_{\odot,H}}),
\end{equation}
where $\mathrm{X_{\odot,Fe}}$ and $\mathrm{X_{\odot,H}}$ refer to the solar abundances of Fe and H. For our work we use $\varphi$ in order to distinguish the planetary abundances and avoid confusion with the stellar values. We often scale the stellar [Fe/H] value for the abundances of the other species in the star. This reduces $\varphi_\mathrm{A,B}$ to
\begin{equation}
    \varphi_\mathrm{A,B} = \log(\mathrm{X_A}/\mathrm{X_B}) - \log(\mathrm{X_{\odot,A}}/\mathrm{X_{\odot,B}}).
\end{equation}
We have now removed the dependence on the measured host star metallicity. As an example, a value of $\varphi_\mathrm{A,B}=2$ means that ratio of species A to B is 100$\times$ that of their solar abundance ratio.

Figure~\ref{fig:chem} (bottom panel) and Table~\ref{tab:chem_fe_ratio_constraints} show the constraints on $\varphi_\mathrm{X,Fe}$ for each of the species in the planets in our survey. The values show that almost every species has an improved error constraint for relative abundances over their absolute values (given in Table~\ref{tab:chem_constraints}), and highlights the strength of HRS to constrain relative abundance ratios. This is an important insight and shows it may be more reliable to use such relative abundance ratios as observational tracers/parameters in retrievals. This also shows that abundance ratios such as the Fe/O or C/O ratio obtained through high-resolution infrared observations may be more reliable than absolute abundances, therefore allowing us to robustly explore formation scenarios \citep{oberg2011, lothringer2021}. 

\begin{figure*}
\centering
	\includegraphics[width=0.99\textwidth,trim={0cm 0cm 0cm 0},clip]{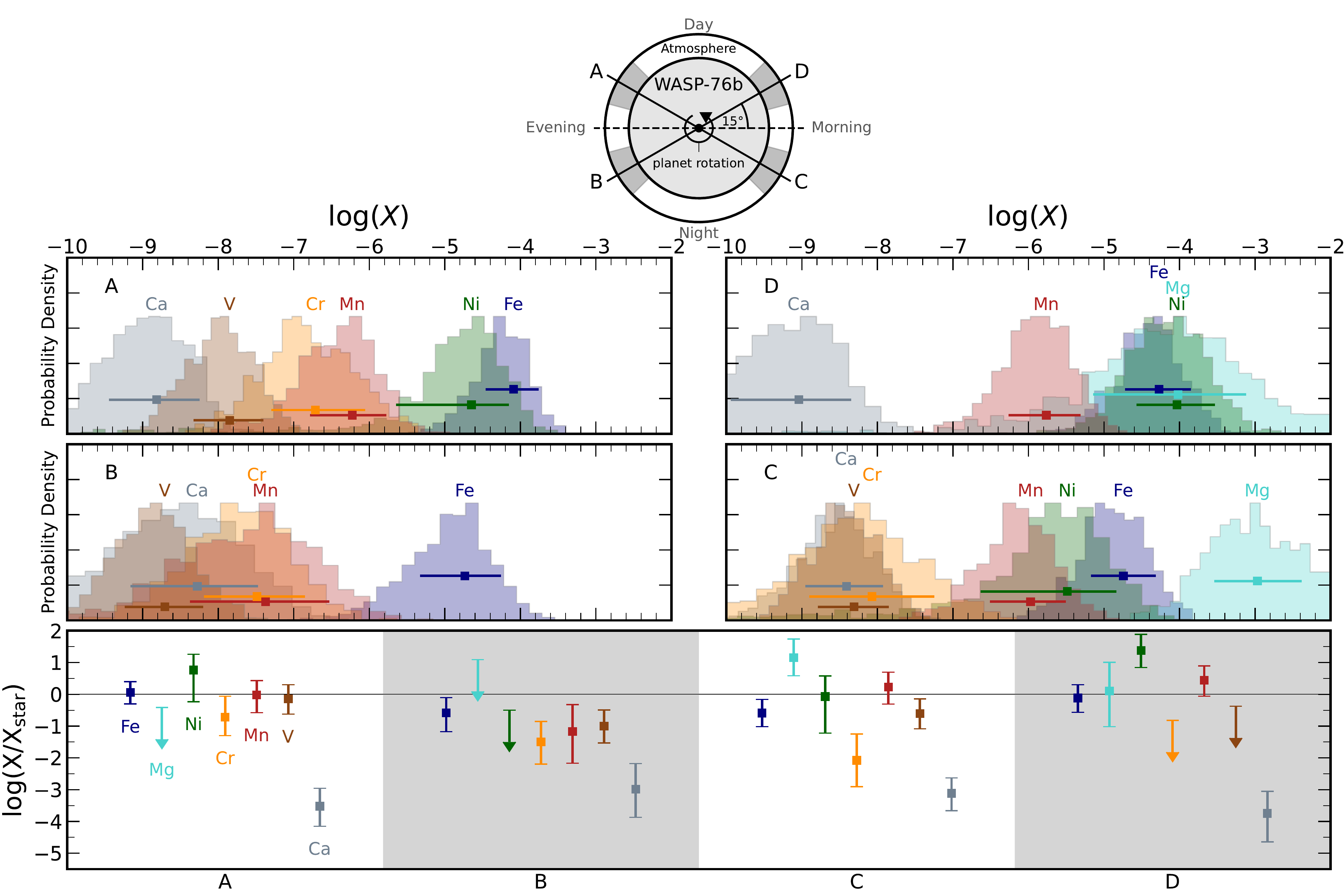}
    \caption{Atmospheric abundance constraints for each species from our HyDRA-2D retrieval of WASP-76~b. The schematic at the top of the figure indicates the regions of the atmosphere which are being probed, with A and D being irradiated by the incident stellar flux. The middle 4 panels show the posterior distributions for each species in each of the regions of the atmosphere, along with their with the median and $\pm1\sigma$ error bars. For clarity we only show those species with a clear posterior peak from the retrieval. The bottom panel shows the ratio of the abundance of each species to the stellar value, with $2\sigma$ upper limits for those species without a clear constraint.}     
\label{fig:wasp76}
\end{figure*}

\begin{figure*}
\centering
	\includegraphics[width=0.99\textwidth,trim={0cm 0cm 0cm 0},clip]{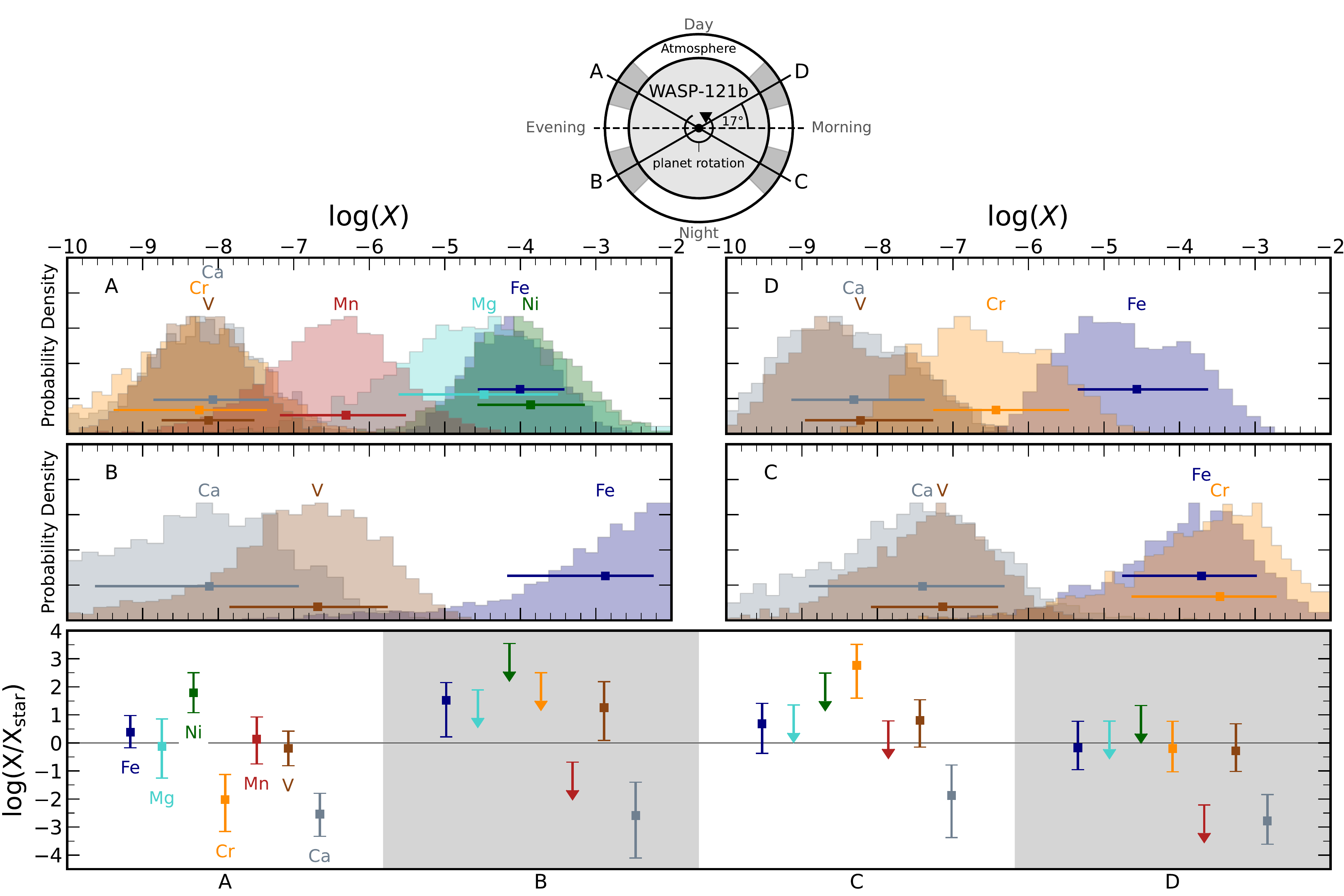}
    \caption{Atmospheric abundance constraints from our HyDRA-2D retrieval of WASP-121~b (see Figure~\ref{fig:wasp76} for details).}     
\label{fig:wasp121}
\end{figure*}

However, we note that relative abundances can also have biased estimates in their values, such as if any RM/CLV affect from the star is greater for one species over another. For each of our planets we chose phase ranges that avoided the stellar contamination to ensure unbiased estimates of the planetary signal which were not influenced by the star. Furthermore, if species have velocity/wavelength shifts relative to one another this can also bias the retrieved ratios, as our retrieval assumes that the line positions of species do not have any relative shifts between one another. These shifts can arise due to species being probed at different altitudes, where wind shear can alter the overall Doppler shifts of spectral signals. For UHJs it is expected that the strongest wind speeds occur at the lowest pressures \citep[e.g.,][]{wardenier2021}, and hence species with strong spectral lines probing these upper regions of the atmosphere will have more significant Doppler shifts.

\subsection{Variation in the terminator of WASP-76~b and WASP-121~b}

Figures~\ref{fig:wasp76} and \ref{fig:wasp121} show the constraints on seven atomic species in the terminator regions of WASP-76~b and WASP-121~b respectively using HyDRA-2D. We did not conclusively detect Ti, TiO, TiH or Na in our spatially-homogeneous and phase-combined retrievals for either planet (see section~\ref{sec:abundances}), and thus they were not included in our HyDRA-2D retrievals for computational efficiency. Our results for both planets are generally consistent with the constraints described above, and most species show similar abundances in each region. This points to a terminator that is broadly homogeneous, with only some slight differences in the chemistry. Hence other factors such as variation in the thermal profile between each region and winds in the upper atmosphere are likely to play a larger role in spectral differences, consistent with previous works in emission geometries \citep{beltz2021}.

The retrievals show that all of the retrieved species show strong constraints in some regions of the atmosphere. In particular, Fe is well constrained in all 4 regions of the terminator for both of the planets, owing to its high abundance and prominent opacity in the optical wavelengths where the ESPRESSO observations probe. We generally find that the abundance is consistent with the stellar value. For WASP-76~b, we find that our abundance constraints for the non-irradiated regions (regions B and C in Figure~\ref{fig:wasp76}) are the weakest with the lowest abundance. This is consistent with previous theoretical works suggesting rain-out of Fe on the night side, clouds obscuring the morning/leading side terminator and/or thermal differences between the two sides leading to a lower and less well constrained abundance \citep{ehrenreich2020, wardenier2021, savel2022, gandhi2022}. On the other hand, we do not see such a difference between the different regions of the terminator for WASP-121~b, which may be due to its higher irradiation preventing significant rain-out/cloud formation. We note that the Fe abundance for WASP-121~b for region B is only a lower limit and is not fully bounded by our retrieval as the posterior extends towards the upper edge of the prior range. However, this is still consistent with the abundance in the other regions.

In addition to Fe, we constrain Mg, Ni, Cr, Mn, V and Ca in both planets, but some species only show upper bounds on the abundance for some regions. For WASP-76~b, Mn shows significant peaks for all regions of the atmosphere, but the Cr and V abundance is only well constrained in regions A, B and C. On the other hand, Ni is only prominently detected in regions A, C and D, with the irradiated regions A and D showing the tightest constraints. One of the most difficult species to constrain on each side was Mg given its fewer spectral lines, and we only obtained clear detections for the morning/leading side of the terminator. Generally however, we find that the regions which only show upper limits are consistent with the values in the constrained regions for all species. 

The abundance constraints for WASP-121~b in Figure~\ref{fig:wasp121} are generally weaker and show a wider uncertainty for each species due to the single night of observations used in our study. While Fe, Ca and V show clear posterior peaks, the other species are not seen in all regions. One of the most intriguing constraints that we observe is for Mn, where we see a clear peak for the species in region A but were not able to clearly constrain it in our spatially-homogeneous retrievals or in any other terminator region of the spatially-resolved and phase-separated retrievals. This may point to some slight offsets in velocity in the Mn signal over the other species, potentially arising from differential wind speeds given that the spectral signals arise from slightly different altitudes for each species. On the other hand, if the Mn is only prominent on one side of the atmosphere, the spatially-homogeneous retrieval may not be able to clearly constrain it given that it will have a stronger blue-shift than species which have more equal prominence across both halves of the terminator. In addition, the upper limit for Mn in region D appears to be inconsistent with the constrained abundance in region A, further supporting this difference in Mn between the two sides. MnS clouds are expected to form at temperatures below $\sim$1100~K (see Table~\ref{tab:cond}), but this is not consistent with our temperature constraints of $\gtrsim$2000~K for the terminator, or with the clear constraint obtained for Mn in WASP-76~b, the planet with the lowest equilibrium temperature in our survey. Hence further observations of the terminator are needed to fully ascertain the variation of Mn in the terminator.

For some species, such as Cr, we find that a higher abundance/stronger constraint in region A results in weaker constraints in region C and vice versa. To separate regions A and C and regions B and D we split the terminator into two halves and retrieve separate chemical abundances for each side. However, abundance constraints for one side may influence the constraints for another, as regions A and C and regions B and D are retrieved simultaneously in the retrieval. Hence, any velocity shifts between species may be compensated by increasing the abundance for one side or the other. In addition, a higher abundance or stronger signal for a given species in one region may spread its spectral features out over the other region of the atmosphere. This is because the spectral lines are broadened and therefore may be partially blended together, thus resulting in a lower abundance and/or a wider uncertainty for the region with weaker absorption. This is similar to that seen for Fe in WASP-76~b, where the more dominant signal from region A increases the uncertainty in region C \citep{gandhi2022}. 

We find that the evening/trailing side for the last part of the transit (region A) generally has the strongest constraints on the abundances for each species. This is because this is the region which is being irradiated and hence has the highest temperatures from our retrievals. The higher temperatures mean that the scale height of the atmosphere is increased and thus these regions of the terminator are the dominant part of the signal. Any east-west meridional circulation patterns would also act to shift the day side hotspot towards the eastern side of the planet and thus closest to region A. Hence we expect this to be the region with the strongest absorption and therefore the most tightly constrained abundances.

\begin{figure*}
\centering
	\includegraphics[width=\textwidth,trim={0cm 0cm 0cm 0},clip]{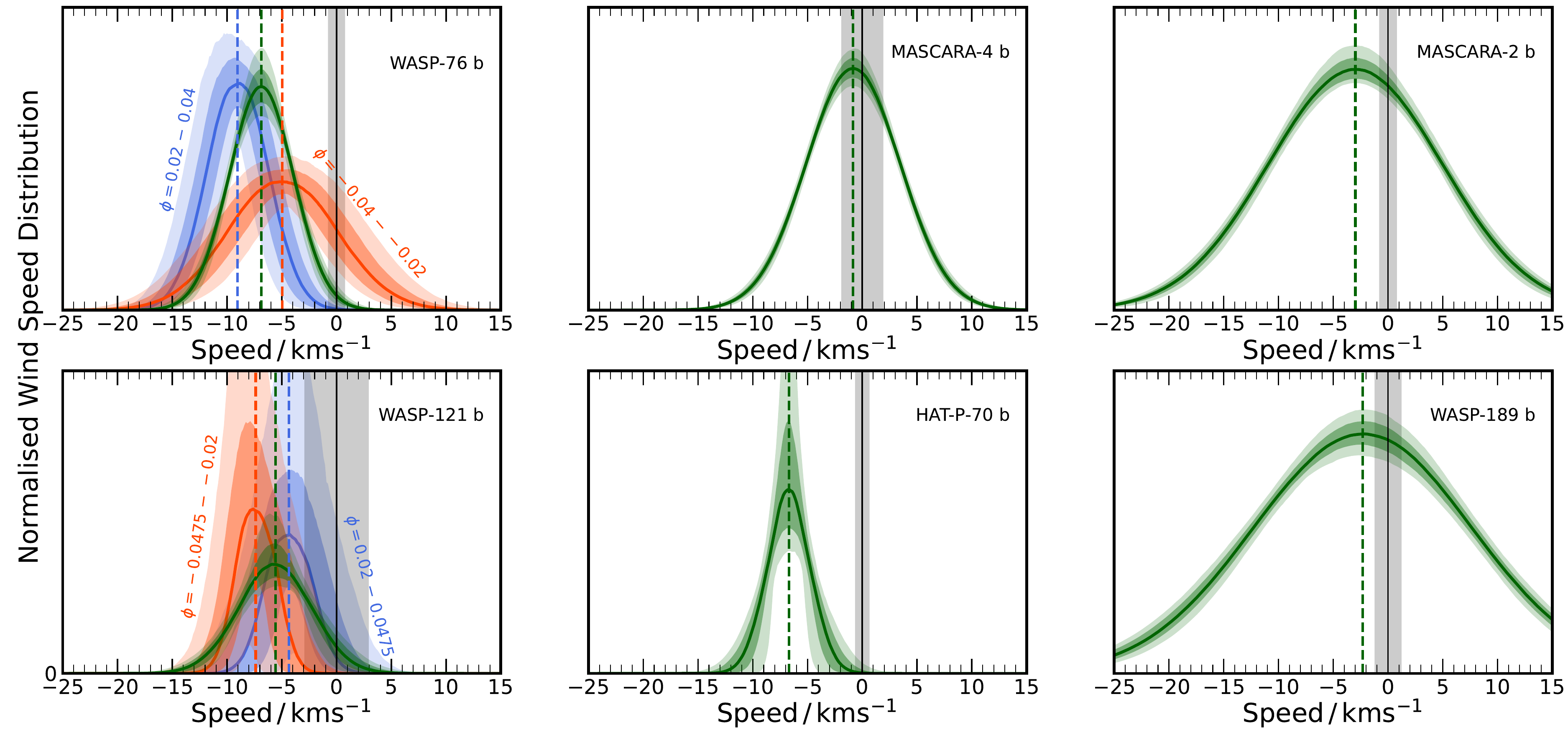}
    \caption{Normalised wind speed distribution in each exoplanet's rest frame, derived from the $\Delta V_\mathrm{sys}$ and $\delta V_\mathrm{wind}$ parameters. The shaded green regions show the 1 and 2$\sigma$ uncertainties on the distribution for the HyDRA-H retrievals (given in Table~\ref{tab:wind_constraints}). We also show the constraints from the HyDRA-2D retrievals of WASP-76~b and WASP-121~b in orange (first part of transit) and blue (last part of transit). The dashed lines indicate the median $\Delta V_\mathrm{sys}$ values. Negative values correspond to a a net blue-shift, as a result of a day-to-night wind. The grey shaded regions show the 3$\sigma$ uncertainty on the known value of $V_\mathrm{sys}$.}     
\label{fig:winds}
\end{figure*}

\subsection{Wind speeds}\label{sec:winds}

Atmospheric winds have a strong influence on the high resolution observations, through Doppler shift and broadening of lines in the observed spectra \citep[e.g.,][]{showman2013}. For such planets, the dominant winds in the upper atmosphere are expected to be those which travel from the day side of the atmosphere towards the night, and thus lead to an overall blue shift of the spectrum as material travels towards the observer \citep[e.g.,][]{kempton2012, flowers2019, wardenier2021, savel2022}. The wind speeds are determined by the balance of day-night temperature contrasts and dissipation mechanisms such as hydrodynamic or magnetic drag \citep[e.g.,][]{gandhi2020_winds, beltz2022}. Figure~\ref{fig:winds} shows the constrained value of $\Delta V_\mathrm{sys}$, the proxy for the wind speed, and the spread in its velocity, $\delta V_\mathrm{wind}$, assuming a uniform day-night wind. The wind broadening kernel is assumed to be a Gaussian profile and is separate from the Doppler shift due to rotation, which is applied as rigid body rotation assuming a tidally locked planet \citep[e.g.,][]{showman2002}. As we account for this planetary rotation, the wind speed distribution in Figure~\ref{fig:winds} is driven by the winds and/or other broadening effects, and therefore can be compared across planets regardless of their rotation. 

The retrieved values of $\Delta V_\mathrm{sys}$ are dependent on the precision of the stellar velocity ($V_\mathrm{sys}$), the time of mid transit (T$_c$) and the orbital period (P). All of the planets are extremely well-studied due to the fact that they orbit bright stars and have short orbital periods, leading to many observed transits over years of baseline. The uncertainties in $V_\mathrm{sys}$ for each star is $\lesssim$0.15~km/s (see Table~\ref{tab:planet_params}). We propagate the uncertainty in T$_c$ and P for each system and find that for each case the change in the measured $\Delta V_\mathrm{sys}$ value is less than 0.3~km/s. We note that these are fitting errors and do not account for the overall systematic uncertainty and may potentially be larger. However, because the retrieved $\Delta K_\mathrm{p}$ and $\Delta V_\mathrm{sys}$ values are correlated \citep[see e.g.,][]{gandhi2022} and the $K_\mathrm{p}$ values have larger uncertainties of 1 to 6 km/s this is actually the largest source of uncertainty on $\Delta V_\mathrm{sys}$. Even with the larger uncertainties on $K_\mathrm{p}$, we still find the total propagated uncertainty in $\Delta V_\mathrm{sys}$ is less than 1~km/s in every case, and cannot explain the overall net blueshifts we observe in the planet atmospheres.

All retrievals for all of the planets have a net blue-shift (i.e. a negative velocity) of the planetary signal relative to the known value of the systemic velocity, which we interpret as evidence of a day-night wind. In addition, the median value of the offset is more than the overall 3$\sigma$ propagated uncertainty in the measured value of $V_\mathrm{sys}$ for every planet except MASCARA-4~b (see Figure~\ref{fig:winds}), and hence cannot solely be explained through errors in the orbital parameters alone. There is a range in values present for the HyDRA-H retrievals, between $-0.84^{+0.17}_{-0.16}$~km/s for MASCARA-4~b to $-6.87^{+0.17}_{-0.17}$~km/s for WASP-76~b, as shown in Table~\ref{tab:wind_constraints}. The constrained blue-shifts are consistent with previous work exploring the Na signatures for a range of transiting hot- and ultra-hot Jupiters \citep{langeveld2022}. From our constraints we do not observe any significant trend in wind velocity across our sample of planets. Hence further observations of the terminator at high resolution are needed to to discern the driving mechanisms of the atmospheric dynamics.

Figure~\ref{fig:winds} also shows the constraints on the wind speeds for WASP-76~b and WASP-121~b for the first and last part of the transit from the HyDRA-2D retrievals. This allows us to compare the wind profiles for each phase region. The results show that, for WASP-76~b, the wind speeds are much higher and the profile has a much smaller spread for the latter part of the transit, in agreement with previous findings \citep{ehrenreich2020, kesseli2021}. Previous retrievals have shown that this could be driven by either the condensation of Fe on the morning/leading side or a high altitude opacity deck which obscures the signal from this part of the atmosphere, with a slight preference for the latter \citep{gandhi2022}. Hence the signal is dominated by the more blue shifted evening/trailing side of the atmosphere and thus the wind speeds are higher with a lower spread in the spectral signal. On the other hand, the retrievals for WASP-121~b indicate a more similar wind speed and $\delta V_\mathrm{wind}$ for the first and last part of the transit. This may be due to the higher irradiation of WASP-121~b, resulting in a hot dayside where the Fe does not rain-out and/or clouds do not form as readily in the photosphere. Our obtained wind speeds are also in agreement with the work by \citet{maguire2023}, who obtained values $\sim - $6~km/s for three separate transits of WASP-121~b.

The value of $\delta V_\mathrm{wind}$, indicative of the spread in the profile, varies significantly across our sample. Three of the planets, MASCARA-4~b, MASCARA-2~b and WASP-189~b, show broader profiles for the wind speed distribution compared to the others. The larger spreads in the profiles of MASCARA-4~b and WASP-189~b could be driven by stellar residuals as these were the two planets in our sample where the RM effect significantly affected the planetary signal. While our RM correction allowed us to recover the planet's signal, it clearly left both positive and negative residuals near the planet's trace, which easily could cause the signal to appear broader. While MASCARA-2~b did not suffer from as much stellar contamination, this could still cause some amount of broadening. Previous studies using this same dataset noted that the Fe I cross correlation profile appeared double peaked \citep[e.g.,][]{stangret2020}, and with our 1D retrieval a double peaked absorption profile could manifest as a large value of $\delta V_\mathrm{wind}$. To fully understand what physical phenomenon is driving the blueshifts and line broadening, further analysis and data are required, which we leave for a subsequent study. For our study, we note that net blueshifts on the order of $\sim$1-7 km/s seem ubiquitous in UHJ atmospheres.

\subsection{Mass constraints for MASCARA-2~b and HAT-P-70~b}\label{sec:masses}

For MASCARA-2~b and HAT-P-70~b only an upper limit has been placed on the planetary mass, and hence we included the surface gravity, log(g), as an additional retrieval parameter for these two planets. The surface gravity of the planet has a strong influence on the strength of spectral features as it determines the scale height of the atmosphere, and previous studies have shown that transit observations can be used to extract the mass of the planet \citep{dewit2013}. For our retrievals of these two planets we use a uniform prior of log($g$) = [2,7] (with $g$ in cm/s$^2$). For a given value of $\log(g)$, the corresponding mass $M$ is
\begin{equation}
\frac{\mathrm{M}}{\mathrm{M_J}} \approx 4.0347\times10^{-4} \times 10^{\log(g)} \times \Bigl(\frac{\mathrm{R_p}}{\mathrm{R_J}}\Bigl)^2,
\end{equation}
where $\mathrm{R_p}$ refers to the radius of the planet, and $\mathrm{M_J}$ and $\mathrm{M_J}$ refer to the radius and mass of Jupiter respectively.

Using our retrieved value of log(g) we derive the masses of $2.16^{+0.21}_{-0.22}$~M$_\mathrm{J}$ for MASCARA-2~b, and $1.66^{+0.20}_{-0.20}$~M$_\mathrm{J}$ for HAT-P-70~b. This is consistent with their upper limits of $<$3.5~M$_\mathrm{J}$ and $<$6.78~M$_\mathrm{J}$ respectively \citep{lund2017,zhou2019}. Hence this method may be viable in providing mass constraints for exoplanets orbiting fast rotating stars such as MASCARA-2~b and HAT-P-70~b in future. We tested the validity of this method by performing retrievals on the WASP-76~b ESPRESSO observations with log(g) as an additional retrieval parameter, and obtained a mass estimate of $0.84^{+0.06}_{-0.08}$~M$_\mathrm{J}$ for the $\phi = +0.02 \,\text{-}\, +0.04$ phase range, consistent with but slightly lower than the measured mass of $0.894^{+0.014}_{-0.013}$~M$_\mathrm{J}$ \citep{ehrenreich2020}.

However, we note that this method of determining the planetary mass has degeneracies with the atmospheric temperature, as both the temperature and mass strongly influence the scale height. The method also relies on the assumption that the mean molecular weight of the atmosphere is determined, which for our case is assumed to be an atmosphere dominant in H and He. The recombination of H to H$_2$ in the atmosphere will reduce the scale height, and may result in the log(g) and hence the mass being overestimated if we assume an H/He rich case. In addition, non-hydrostatic effects in the upper atmosphere can also influence the overall extent of spectral features and hence bias the determined planet mass. Therefore, further independent verification of the masses of these two objects is needed to robustly test these biases further.

\section{Conclusions}\label{sec:conc}

We retrieve the chemical abundances of eleven neutral atomic and molecular species in six UHJs through optical observations with the ESPRESSO and HARPS-N/HARPS spectrographs. We use the HyDRA-H and HyDRA-2D retrieval frameworks to perform retrievals of the terminator observations \citep{gandhi2019_hydrah, gandhi2022}. From our retrieval survey we find that:

\begin{itemize}
    \item All of the planets show clear constraints on the atmospheric abundance of Fe given its high opacity and abundance, with values consistent with the stellar metallicity across the sample. This points to UHJs having at or near stellar abundances for refractory species, consistent with previous high-resolution studies \citep{pelletier2023} and previous work exploring alkali metals at low-resolution \citep{welbanks2019}.
    
    \item In addition to Fe, we also retrieve for the abundances of Mg, Ni, Cr, Mn, V, Ca, Ti, TiO, TiH and Na for each planet. Our abundance constraints for Mg, Ni, Cr, Mn and V show a much greater variation across the sample of UHJs, with generally wider uncertainties or upper limits given their lower abundance and/or opacity. This highlights the challenge of using the abundance of a single species as a proxy for atmospheric metallicity, as these planets show significant compositional diversity.
    
    \item Three species in our sample, Ca, Ti and Na, show constraints/upper limits which are below that expected if each species was present at stellar metallicity for all of the planets. However, chemical equilibrium models provide a good explanation as these are the species which are some of the most strongly ionised at high temperature and low pressure. Hence the abundance of neutral species containing these atoms is reduced. For Ca and Ti the retrievals do still indicate even lower abundances than suggested by chemical equilibrium, which may be caused by strong vertical chemical gradients for these species and/or the presence of CaTiO$_3$ clouds quenching these two species below the photosphere. We are able to constrain Ti and TiO for only WASP-189~b, the hottest planet in our sample, supporting the hypothesis of Ti being quenched on the night sides of the other cooler planets.
       
    \item We find that abundance ratios between species show tighter error bars than the absolute values for almost all species and planets, consistent with previous works \citep[e.g.,][]{gibson2022, maguire2023, pelletier2023}. This near universal finding across our sample indicates that HRS has an inherent strength in constraining abundance ratios of species in exoplanet atmospheres. This arises because HRS is more sensitive to the relative line strengths between species given that the contribution from the spectral continuum is often removed/reduced during the analysis. This highlights the potential of using HRS for high precision abundance ratio measurements, such as for refractory/volatile or C/O ratios. In addition, using low resolution observations such as with HST and JWST could be key in further constraining the abundances of trace species when used in tandem with HRS \citep{gandhi2019_hydrah}.
    
    \item For two of the planets, WASP-76~b and WASP-121~b, we perform spatially-resolved and separated-phase retrievals to explore the variation of each species in each region of the terminator. These retrievals show that the chemical abundances of each species at the terminator does not vary significantly and hence that there is no significant chemical asymmetry for most of the species. For both planets we are able to retrieve Fe, Mg, Ni, Cr, Mn, V and Ca in at least one part of the atmosphere, with the regions where they were not conclusively constrained generally showing upper limits consistent to the retrieved abundances in the detected regions. For WASP-76~b we did constrain a lower abundance for Fe with a wider uncertainty for the less irradiated regions of the terminator, in agreement with previous work indicating that rain-out, cloud formation and/or thermal differences between the leading and trailing limbs may be present \citep{ehrenreich2020, wardenier2021, savel2022, gandhi2022}.
    
    \item In addition to the chemical abundances, we explored the atmospheric winds present on these planets through the wavelength shift and broadening of the planetary spectrum compared with the known orbital parameters. This Doppler shift arises as a result of winds in the upper atmosphere transporting material from the day side to the night side. From our retrievals we find blue-shifts with median values $\sim2.3-9$~km/s, but with no significant trend across the sample of planets in our survey. For WASP-76~b, we constrain the highest wind speed for the end of the transit of $9.03^{+1.03}_{-1.01}$~km/s, but the first quarter of the transit shows a much lower $4.94^{+1.62}_{-1.55}$~km/s velocity offset. This is consistent with previous observations and analyses \citep{ehrenreich2020, kesseli2021, wardenier2021, gandhi2022}.
    
    \item We derive the masses of MASCARA-2~b and HAT-P-70~b by including a free parameter for the log(g) of the planets, constraining masses of $2.16^{+0.21}_{-0.22}$~M$_\mathrm{J}$ and $1.66^{+0.20}_{-0.20}$~M$_\mathrm{J}$ respectively. This is consistent with their upper limits in Table~\ref{tab:planet_params}. This presents a potential way to determine the masses of planets around fast rotating stars, but we note that various degeneracies may exist in using the spectroscopic mass, and further work is needed to fully explore these biases.
\end{itemize}

Future work can expand the sample across a wider range of equilibrium temperatures and planet masses/radii in order to determine whether these trends vary across planetary system properties. Exoplanets with slightly cooler equilibrium temperatures nearer to 2000~K are of particular interest as many refractory species are expected to condense out of the atmosphere and the transition between ultra-hot Jupiters and hot Jupiters occurs \citep[e.g.,][]{fortney2008}. Our retrievals could also be expanded to include ionic species in the atmosphere, as the chemical equilibrium models have shown that some species are significantly ionised at temperatures above $\sim$2500~K. However, the ionisation of atomic species is often a very strong function of pressure and may therefore require an abundance profile that varies with atmospheric altitude instead of our assumption of vertically homogeneous chemistry. The ionic species also often probe the exosphere, which can make inferring abundances more challenging \citep[e.g.,][]{zhang2022, maguire2023}. In addition, we can also explore trends across the sample of planets in emission spectroscopy, which probe the hotter day side and are more sensitive to the thermal profile of the atmosphere \citep[e.g.,][]{line2021, cont2022, brogi2023, vansluijs2023}.

Abundance constraints for the refractory species are of particular importance as they are expected to be accreted onto the planet as solids given their high sublimation temperatures, and hence provide a measure of the accreted rock during formation \citep{lothringer2021, knierim2022, chachan2022}. Future work can combine these optical observations with those in the infrared such as JWST and ground-based observatories such as IGRINS/Gemini-S and CRIRES+/VLT which can constrain volatile species. Such studies will provide important insights into the refractory-to-volatile ratios of exoplanets, allowing us to compare with carbon and oxygen bearing species to determine formation and migration scenarios \citep[e.g.,][]{oberg2011, mordasini2016}. For such retrievals vertically homogeneous chemistry assumptions may also need to be relaxed as abundances of species such as H$_2$O are strong functions of pressure \citep{parmentier2018}, and can otherwise lead to biased retrieved estimates \citep{pluriel2022}.

This work demonstrates the similarities and differences in the chemical abundances in our sample of ultra-hot Jupiters. Furthermore, the high-resolution observations are sensitive to atmospheric winds which transport thermal energy and material between the day and night side, resulting in a net blue-shift of the spectrum, which we are able to clearly constrain for the planets in our sample. High-resolution spectroscopy will therefore play a key role in exploring atmospheric chemistry and dynamics on exoplanets in upcoming years, in particular with the next generation of large ground-based facilities such as ELT \citep{maiolino2013}.


\section*{Acknowledgments}
SG is grateful to Leiden Observatory at Leiden University for the award of the Oort Fellowship. This work was performed using the compute resources from the Academic Leiden Interdisciplinary Cluster Environment (ALICE) provided by Leiden University. We also utilise the Avon HPC cluster managed by the Scientific Computing Research Technology Platform (SCRTP) at the University of Warwick. 
YZ and IS acknowledge funding from the European Research Council (ERC) under the European Union's Horizon 2020 research and innovation program under grant agreement No. 694513. M.B.\ acknowledges support from the STFC research grant ST/T000406/1. NPG and CM gratefully acknowledge support from Science Foundation Ireland and the Royal Society via a University Research Fellowship and Enhancement Award. We thank the anonymous referee for a careful review of our manuscript.

%

\vspace{5mm}
\facilities{VLT (ESPRESSO), TNG (HARPS-N), La Silla (HARPS)}





\appendix

\section{Condensation Temperatures for Refractory Species}

\begin{table*}[h]
    \centering
    \def\arraystretch{1.5}
\begin{tabular}{c|c|c|c|c}
\textbf{Species} & \textbf{Condensate} & \textbf{P-T relation} & \textbf{T(0.1mbar)/K} & \textbf{Reference} \\
\hline
Ca/Ti & CaTiO$_3$ & $\frac{10^4}{\mathrm{T}} = 5.125 - 0.277\log(\mathrm{P}) - 0.554[\mathrm{Fe/H}]$ &  1604 & \citet{wakeford2017} \\
Fe & Fe           & $\frac{10^4}{\mathrm{T}} = 5.44 - 0.48\log(\mathrm{P}) - 0.48[\mathrm{Fe/H}]$ &  1359 & \citet{visscher2010} \\
Mg &Mg$_2$SiO$_4$ & $\frac{10^4}{\mathrm{T}} = 5.89 - 0.37\log(\mathrm{P}) - 0.73[\mathrm{Fe/H}]$ & 1357 & \citet{visscher2010} \\
Mg & MgSiO$_3$    & $\frac{10^4}{\mathrm{T}} = 6.26 - 0.35\log(\mathrm{P}) - 0.70[\mathrm{Fe/H}]$ & 1305 & \citet{visscher2010} \\
Cr & Cr           & $\frac{10^4}{\mathrm{T}} = 6.576 - 0.486\log(\mathrm{P}) - 0.486[\mathrm{Fe/H}]$ & 1174 & \citet{morley2012} \\
Mn & MnS          & $\frac{10^4}{\mathrm{T}} = 7.45 - 0.42\log(\mathrm{P}) - 0.84[\mathrm{Fe/H}]$ & 1095 & \citet{visscher2006} \\
Na & Na$_2$S      & $\frac{10^4}{\mathrm{T}} = 10.05 - 0.72\log(\mathrm{P}) - 1.08[\mathrm{Fe/H}]$ & 774 & \citet{visscher2006} \\

\end{tabular}
    \caption{Condensation temperatures for the refractory species in our work. The third column shows the P-T relation of the condensation curve, with the temperature T in K, the pressure P in bar and the metallicity of the atmosphere [Fe/H].}
    \label{tab:cond}
\end{table*}

\clearpage

\section{Temperature Profiles}

\begin{figure*}[h]
\centering
	\includegraphics[width=0.72\textwidth,trim={0cm 0cm 0cm 0},clip]{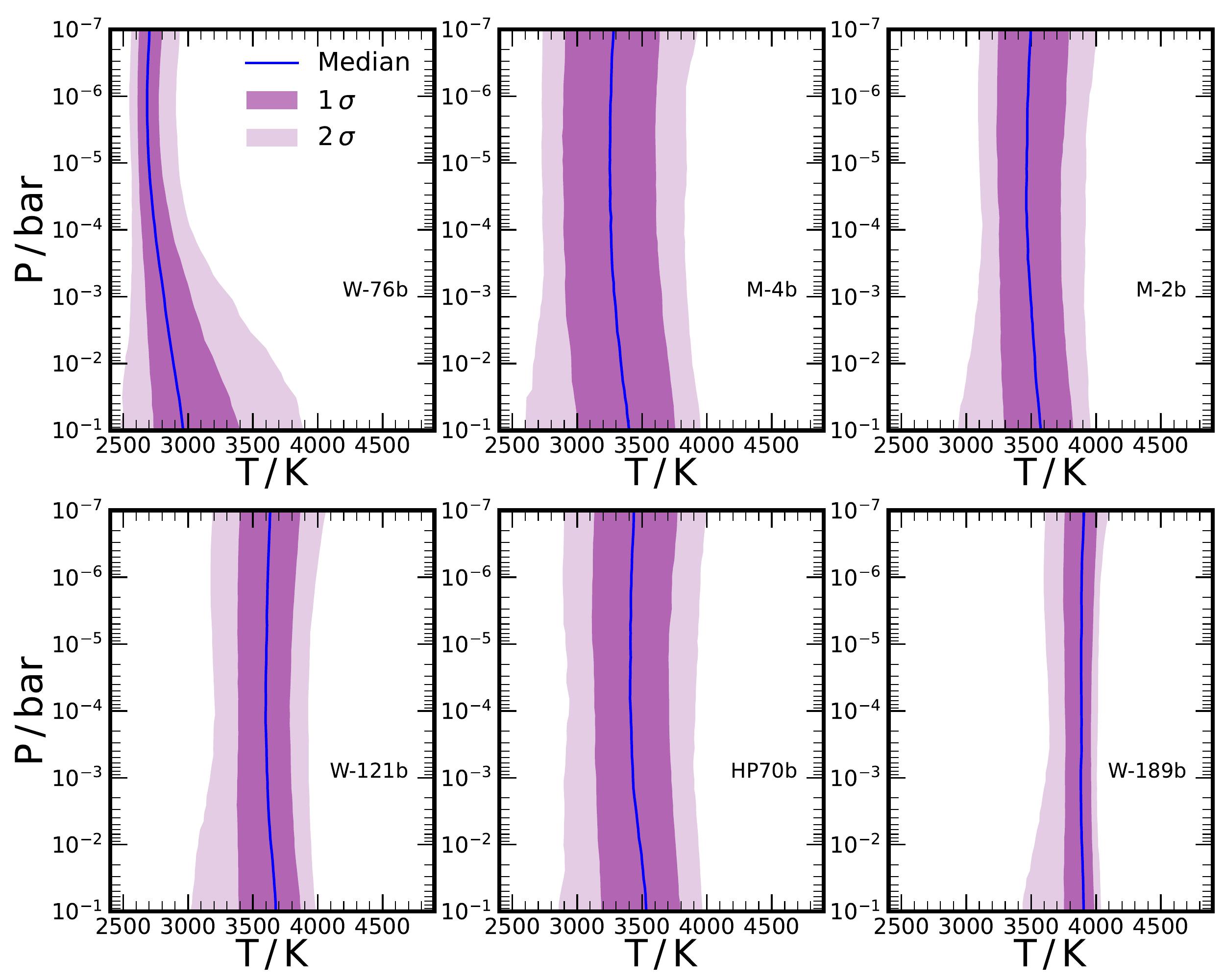}
    \caption{Retrieved temperature profiles for the HyDRA-H retrievals for each of the planets in our survey. The median profile is given by the blue line, with the darker and lighter shades of purple indicating the 1 and 2$\sigma$ confidence ranges respectively.}    
\label{fig:PT}
\end{figure*}


\section{Relative abundances}

\begin{table*}[h]
   \centering
    \def\arraystretch{1.5}
\begin{tabular}{c|cccccc}
Species & \multicolumn{1}{c}{\bf{WASP-76~b}} & \multicolumn{1}{c}{\bf{MASCARA-4~b}} & \multicolumn{1}{c}{\bf{MASCARA-2~b}} & \multicolumn{1}{c}{\bf{WASP-121~b}} & \multicolumn{1}{c}{\bf{HAT-P-70~b}} & \multicolumn{1}{c}{\bf{WASP-189~b}}\\
\hline
Mg & $0.57^{+0.33}_{-0.37}$ & $0.33^{+0.16}_{-0.15}$ & $0.27^{+0.18}_{-0.20}$ & $-1.35^{+0.66}_{-0.90}$ & $0.74^{+0.31}_{-0.33}$ & $0.24^{+0.22}_{-0.24}$\\
Ni & $0.69^{+0.17}_{-0.17}$ & $-0.16^{+0.21}_{-0.25}$ & $<-0.08$ & $0.85^{+0.27}_{-0.28}$ & $<0.660$ & $0.57^{+0.20}_{-0.22}$\\
Cr & $-1.16^{+0.16}_{-0.19}$ & $-2.81^{+0.23}_{-0.27}$ & $-1.28^{+0.21}_{-0.24}$ & $-0.71^{+0.26}_{-0.26}$ & $-1.25^{+0.36}_{-0.41}$ & $-0.42^{+0.14}_{-0.15}$\\
Mn & $0.36^{+0.13}_{-0.14}$ & $-1.82^{+0.28}_{-0.32}$ & $-0.63^{+0.25}_{-0.28}$ & $<-1.78$ & $0.36^{+0.32}_{-0.35}$ & $-0.74^{+0.31}_{-0.34}$\\
V & $-0.57^{+0.14}_{-0.15}$ & $-2.06^{+0.20}_{-0.24}$ & $<-2.70$ & $-0.24^{+0.17}_{-0.17}$ & $-0.65^{+0.25}_{-0.27}$ & $-1.69^{+0.40}_{-0.46}$\\
Ca & $-3.14^{+0.19}_{-0.22}$ & $-2.81^{+0.14}_{-0.16}$ & $-2.45^{+0.17}_{-0.19}$ & $-2.74^{+0.24}_{-0.25}$ & $-2.86^{+0.31}_{-0.38}$ & $-2.23^{+0.15}_{-0.16}$\\
$^*$Ti & $<-2.66$ & $<-2.81$ & $<-1.67$ & $<-2.16$ & $<-2.13$ & $-1.48^{+0.13}_{-0.16}$\\
Na & $<-1.63$ & $<-2.61$ & $<-2.00$ & $<-1.52$ & $<-2.02$ & $<-2.27$\\
\end{tabular}
    \caption{Values of $\varphi_\mathrm{X,Fe}$ and 1$\sigma$ errors for each species X for the planets in our survey. Where no significant constraints were retrieved, we show the 2$\sigma$ upper limit. $^*$Ti value is the sum of the constraints for atomic Ti, and molecular TiO and TiH.}
    \label{tab:chem_fe_ratio_constraints}
\end{table*}

\clearpage

\section{Constraints on Winds}

\begin{table*}[h]
    \centering
    \def\arraystretch{1.5}
\begin{tabular}{c|lc|cc}
\bf{Planet} && \bf{$\mathbf{\phi}$ range} & \bf{$\mathbf{\Delta V_\mathrm{sys}}$/ km~s$^{-1}$} & \bf{$\mathbf{\delta V_\mathrm{wind}}$/km~s$^{-1}$} \\
\hline
\bf{WASP-76~b}  && $-0.04 \,\text{-}\, -0.02$, $+0.02 \,\text{-}\, +0.04$    & $-6.87^{+0.17}_{-0.17}$ & $6.93^{+0.55}_{-0.5}$\\
\cline{2-5}
                & 2D &$-0.04 \,\text{-}\, -0.02$     & $-4.94^{+1.62}_{-1.55}$ & $11.58^{+0.97}_{-0.93}$\\
                & 2D & $+0.02 \,\text{-}\, +0.04$    & $-9.03^{+1.03}_{-1.01}$ & $6.46^{+0.55}_{-0.56}$\\
\hline
\bf{MASCARA-4~b}  && $-0.03 \,\text{-}\, -0.015$, $+0.00 \,\text{-}\, +0.01$   & $-0.84^{+0.17}_{-0.16}$ & $10.14^{+0.4}_{-0.43}$\\
\hline
\bf{MASCARA-2~b}  && $-0.022 \,\text{-}\, -0.005$, $+0.005 \,\text{-}\, +0.022$  & $-2.97^{+0.32}_{-0.36}$ & $18.75^{+0.76}_{-0.86}$\\
\hline
\bf{WASP-121~b}  && $-0.0475 \,\text{-}\, -0.02$, $+0.02 \,\text{-}\, +0.0475$  & $-5.58^{+0.4}_{-0.44}$ & $7.92^{+1.1}_{-1.31}$\\
\cline{2-5}
                 & 2D & $-0.0475 \,\text{-}\, -0.02$  & $-7.37^{+1.46}_{-1.34}$ & $4.02^{+1.39}_{-1.46}$\\
                 & 2D & $+0.02 \,\text{-}\, +0.0475$  & $-4.33^{+1.72}_{-1.57}$ & $4.75^{+1.52}_{-1.64}$\\
\hline
\bf{HAT-P-70~b}  && $-0.01 \,\text{-}\, +0.025$   & $-6.7^{+0.29}_{-0.29}$ & $4.59^{+1.23}_{-1.28}$\\
\hline
\bf{WASP-189~b}  && $-0.033 \,\text{-}\, +0.02$   & $-2.33^{+0.5}_{-0.5}$ & $23.59^{+1.1}_{-1.19}$\\

\end{tabular}
    \caption{Phase ranges and wind constraints for each of the planets in our survey. The right 2 columns give the constraints on $\Delta V_\mathrm{sys}$ and $\delta V_\mathrm{wind}$ and their 1$\sigma$ errors. For WASP-76~b and WASP-121~b, we also show the constraints from the HyDRA-2D retrievals on each of the separate phase ranges.}
    \label{tab:wind_constraints}
\end{table*}

\bibliography{refs}{}
\bibliographystyle{aasjournal}



\end{document}